\documentclass[twocolumn]{aastex63}
\usepackage{comment}
\usepackage{color,longtable}

\newcommand{\fn}[1]{\textcolor{black}{#1}}

\received{}
\revised{}
\accepted{\today}
\graphicspath{{./}{figures/}}
\usepackage{lineno}

\begin{document}
\title{
Unveiling Stellar Feedback and Cloud Structure in the $\rho$ Ophiuchi A Region with ALMA and JWST: Discovery of Substellar Cores, C$^{18}$O Striations, and Protostellar Outflows}

\email{fumitaka.nakamura@nao.ac.jp}

\author[0000-0001-5431-2294]{Fumitaka Nakamura}
\affiliation{National Astronomical Observatory of Japan, Mitaka, Tokyo 181-8588, Japan}
\affiliation{Department of Astronomy, The University of Tokyo, 7-3-1, Hongo, Bunkyo-ku, Tokyo, 113-0033, Japan}
\affiliation{The Graduate University for Advanced Studies (SOKENDAI), 
2-21-1 Osawa, Mitaka, Tokyo 181-8588, Japan}

\author[0000-0002-8049-7525]{Ryohei Kawabe}
\affiliation{National Astronomical Observatory of Japan, Mitaka, Tokyo 181-8588, Japan}

\author[0009-0006-1731-6927]{Shuo Huang}
\affiliation{National Astronomical Observatory of Japan, Osawa 2-21-1, Mitaka, Tokyo 181-8588, Japan}
\affiliation{Department of Physics, Graduate School of Science, Nagoya University, Furocho, Chikusa,
Nagoya 464-8602, Japan}

\author[0000-0003-1549-6435]{Kazuya Saigo}
\affiliation{Department of Physics and Astronomy, Graduate School of Science and Engineering, Kagoshima University, \\
1-21-35 Korimoto, Kagoshima, Kagoshima 890-0065, Japan}

\author[0000-0001-9304-7884]{Naomi Hirano}
\affiliation{
Institute of Astronomy and Astrophysics, Academia Sinica \\
11F of Astronomy-Mathematics Building, AS/NTU
No.1, Sec. 4, Roosevelt Rd, Taipei 106319, Taiwan, R.O.C.}

\author[0000-0003-0845-128X]{Shigehisa Takakuwa}
\affiliation{Department of Physics and Astronomy, Graduate School of Science and Engineering, Kagoshima University, \\
1-21-35 Korimoto, Kagoshima, Kagoshima 890-0065, Japan}

\author[0000-0002-2067-629X]{Takeshi Kamazaki}
\affiliation{National Astronomical Observatory of Japan, Mitaka, Tokyo 181-8588, Japan}


\author[0000-0002-6510-0681]{Motohide Tamura}
\affiliation{National Institutes of Natural Sciences, Astrobiology Center, 2-21-1, Osawa, Mitaka, Tokyo 181-8588, Japan}
\affiliation{ Department of Astronomy, The University of Tokyo, 7-3-1, Hongo, Bunkyo-ku, Tokyo, 113-0033, Japan}

\author[0000-0002-9289-2450]{James Di Francesco}
\affiliation{NRC Herzberg Inst of Astropfhysics, 
5071 W Saanich Rd, Victoria BC V9E 2E7, British Columbia, Canada}

\author[0000-0001-7594-8128]{Rachel Friesen}
\affiliation{David A. Dunlap Department of Astronomy \& Astrophysics, University of Toronto, 50 St. George St., Toronto ON, Canada}

\author[0000-0002-2707-7548]{Kazunari Iwasaki}
\affiliation{National Astronomical Observatory of Japan, Mitaka, Tokyo 181-8588, Japan}
\affiliation{The Graduate University for Advanced Studies (SOKENDAI), 
2-21-1 Osawa, Mitaka, Tokyo 181-8588, Japan}

\author{Chihomi Hara}
\affiliation{NEC Corporation Fuchu, Tokyo 183-8501, Japan}
\affiliation{Department of Astronomy, The University of Tokyo, 7-3-1, Hongo, Bunkyo-ku, Tokyo, 113-0033, Japan}

\begin{abstract}
In clustered star-forming regions, stellar feedback—such as H{\sc ii} regions/photon-dominated regions (PDRs), and protostellar jets/outflows—shapes cloud structures and influences star formation. Using high-resolution ALMA millimeter and JWST infrared data, we analyze the cloud structure and the impact of stellar feedback in the \fn{nearest} dense cluster-forming region Oph A.
\fn{All 6 known Class 0/I and 2 of 6 Flat Spectrum/Class II objects are detected in the 1.3 mm dust continuum. Additionally, we newly detected 7 substellar cores, three of which show compact near-infrared emission, suggesting they are young substellar objects. The remaining cores, with 
masses of $\sim 10^{-2}$ M$_\odot$ and mean densities of $\sim 10^8$ cm$^{-3}$, are likely gravitationally bound. They appear connected by faint CO finger-like structures extending from the triple Class 0 system VLA1623-2417 Aa+Ab+B, suggesting they may have been ejected from the close binary VLA1623 Aa+Ab.}
$^{12}$CO and near-infrared data reveal multiple protostellar outflows. From the comparison, we identified \fn{several new outflows/jets}, and shocked structures associated to the GSS30 large bipolar bubble. Strong $^{12}$CO emission traces the eastern edge of the Oph A ridge, forming part of the expanding H{\sc ii}/PDR bubble driven by the nearby Herbig Be star S1. 
The northern ridge appears \fn{``blown out”}, with warm gas flowing toward GSS 30, injecting additional \fn{turbulent momentum}. Several C$^{18}$O striations in the S1 bubble align with magnetic fields, and position-velocity diagrams show wave-like patterns, possibly reflecting magnetohydrodynamic waves. Stellar feedback significantly influences Oph A’s cloud structure.
\end{abstract}


\section{Introduction}
\label{sec:intro}

Most stars form in clustered environments \citep{lada03}, 
where cloud structures and future star formation are significantly influenced by the ongoing star formation activity \citep{elmegreen77, 
 maclow04,mckee07,ballesteros20}. While protostars originate from compact dense cores, typically $\sim$ 0.05 pc or less in size, the outflows they eject extend their influence across much larger scales, affecting the surrounding gas over distances of up to $\sim$1 pc \citep[e.g.,][]{bally16}. This phenomenon therefore introduces significant complexity to a star-forming cloud's structure and kinematics \citep{krumholz14, nakamura14b}. 
In addition, radiation from intermediate-mass and high-mass stars can influence even larger volumes of their parent clouds, dissociating interstellar molecules and ionizing the surrounding gas \citep{krumholz14}. These outflows and stellar radiation not only disrupt the surrounding material but also regulate the mass of forming stars by dynamically dispersing gas \citep{matzner00}. By transferring significant momentum to the ambient gas, outflows also help maintain cloud turbulence, which is crucial for shaping both the cloud's structure and the process of star formation \citep{li06,matzner07,offner17}. To understand fully the complex nature of star formation in clustered environments, it is therefore essential to study how stellar feedback shapes associated cloud structures.

Since protostars are deeply embedded in the dense regions of molecular clouds, both millimeter/submillimeter and infrared  observations can offer complementary insights for star formation studies. Millimeter/submillimeter instruments, like ALMA, excel at tracing the distribution and kinematic properties of cold, dense gas, protostellar envelopes, and molecular outflows from young stars \citep[e.g.,][]{andre93,kamazaki01,difrancesco04,friesen10,kawabe18,kamazaki19,hara21}. In contrast, near-infrared observations are ideal for detecting young stars and protostars hidden within these dense cloud regions, as well as identifying energetic protostellar jets \citep[e.g.,][]{bontemps01, khanzadyan04, zhang09,evans09}. Recently, JWST has provided stunning near-infrared images of star-forming regions, revealing complex cloud structures and embedded protostars in remarkable detail \citep{pontoppidan22}.

The $\rho$ Ophiuchi ($\rho$ Oph) region \citep{wilking08}, located at a distance of 138 pc \citep{lombardi08,ortiz17a,  zucker20}, is one of the nearest and most active star cluster-forming clouds. It has been extensively studied across multiple wavelengths over the past decades \citep[e.g.,][]{loren86, motte98, bontemps01,  kamazaki01, difrancesco04, maruta10, ladjelate20}. Millimeter observations have revealed that the $\rho$ Oph cloud complex contains several dense substructures, including Oph A, Oph B1, Oph B2, Oph C, and others \citep{loren86}. 
Protostars at various evolutionary stages—Class 0, I, and II—are distributed throughout the region \citep{andre90,andre93,motte98}.
An important characteristic of the $\rho$ Oph region is the apparent influence of stellar feedback from nearby high-mass and intermediate-mass stars in the Sco OB2 Association on the entire cloud complex. Some studies suggest that star formation within the cloud complex may have been triggered by this feedback \citep{loren86}. In certain areas, a linear alignment of dense subclumps and cores is observed \citep{motte98, kamazaki19, ladjelate20}, further hinting at the role of external forces in shaping the region's structure.

Many millimeter/submillimeter 
 previous studies have focused on the $\rho$ Oph's subclump, Oph A \citep{andre93, difrancesco04, bourke12, nakamura12b, friesen14,chen18b}. Oph A is elongated from north to south, and in the densest part of the ridge, several protostellar and prestellar cores have been identified \citep{difrancesco04, friesen14, kawabe18,friesen18,friesen24}. The region also contains the prototypical Class 0 object, VLA 1623, located in its southwestern part, which drives a powerful, collimated protostellar outflow \citep{andre90}. The ALMA observations of the VLA 1623 outflow were recently presented by \citet{hara21}.
\citet{kawabe18} reported that the densest core in Oph A is an extremely-young Class 0 object, 
shortly after the formation of the first hydrostatic core \citep[see also][]{kirk17b,friesen18}. Interestingly, this core is associated with X-ray activity, suggesting that magnetic activity has already begun in this very early stage of evolution.
In this paper, through a comparison of ALMA and JWST images of the Oph A region, we aim to uncover the detailed cloud structures of Oph A, with a particular focus on stellar feedback.

The paper is organized as follows: We first describe the details of the ALMA and JWST observations in Section \ref{sec:obs}. In Section \ref{sec:result}, we present the ALMA and JWST images of Oph A and discuss the cloud structures. 
Section \ref{sec:comparison} focuses on detailed comparison between the ALMA and JWST images and attempt to reveal complex cloud structures, focusing on the impact of stellar feedback.
Finally, Section \ref{sec:summary} concludes with a summary of our results.

\begin{table}
 \centering
  \caption{Parameters of the ALMA observations toward Oph A$^a$}
  \begin{tabular}{lcc}
 \hline \hline 
 Parameters &  12-m & 7-m  \\
 \hline
 Target &  Oph A & Oph A  \\
 Observation date &  2015 May 1  & 2014 Aug. 14--17 \\
 Number of pointings & 150 & 53 \\
Number of antennas & 40 & 10  \\
Minimum baseline & 12.5 $k\lambda$ & 8.1 $k\lambda$  \\
Maximum baseline &  348 $k\lambda$ & 48 $k\lambda$  \\
On-Source Time &  33 min & 50 min \\
Bandpass Calibrator &  J1517-2422 & J1733-1304  \\
Phase Calibrator &  J1625-2527 & J1625-2527 \\
Absolute Flux Calibrator &  Titan & Mars \\
 \hline 
\end{tabular}
  \label{tab:obs}
\tablecomments{
Project code 2013.1.00839.S
}
\end{table}

\begin{table*}
  \caption{Target continuum and lines of the ALMA observations}
  \begin{tabular}{ccccc}
 \hline \hline 
 Parameters & 1.3 mm & $^{12}$CO (J=2--1) & $^{13}$CO (J=2–-1) & C$^{18}$O (J=2--1) \\
 \hline
 Frequency (GHz) & 226 (218 and 234) & 230.538000 & 220.398684 & 219.560358 \\
Bandwidth/Channel width & 7 GHz  & 0.2 km s$^{-1}$ & 0.2 km s$^{-1}$ & 0.2 km s$^{-1}$ \\
beam size & 1$\farcs$41$\times$ 0$\farcs$90  & 
1$\farcs$41$\times$ 0$\farcs$91 & 
1$\farcs$47$\times$ 0$\farcs$94  & 
1$\farcs$48$\times$ 0$\farcs$95 \\
beam PA (deg.) & --89.4  & 
--89.5& --88.8 & --89.2\\
noise level$^a$ & 0.25 mJy beam$^{-1}$ & 
42 mJy beam$^{-1}$  & 
55 mJy beam$^{-1}$& 
36 mJy beam$^{-1}$ \\
 \hline 
\end{tabular}
  \label{tab:obs2}
  \tablecomments{$^a$ rms per channel for line data.  \\
  All the data are combined 12 m and 7 m Array data. See  \citet{kawabe18} and \citet{hara21} for details.
  In the present paper, we use both the original images having resolution of $\sim 1$\arcsec \  and the images smoothed to 2\arcsec \ resolution.
  \fn{The noise level of the smoothed continuum image is 0.445 mJy beam$^{-1}$}.}
\end{table*}

\section{Observations and Data}
\label{sec:obs}

\subsection{ALMA Observations}

We conducted wide-field mosaic observations of the Oph A region using the ALMA 12-m Array and the 7-m Array (ACA) in Band 6 during Cycle 2 (PI: F. Nakamura, Project code 2013.1.00839.S). The reference position for both arrays was set to ($\alpha_{J2000.0}$, $\delta_{J2000.0}$) = (16$^h$27$^m$26\arcsec.507, --24$^\circ$31\arcmin28\arcsec.63), which is the same position  used in the Oph B2 observations \citep{kamazaki19}
as they were executed as a part of the common project.

The observed field covers an area of 160\arcsec $\times$ 190\arcsec \, in Oph A. We used 150 pointings for the 12 m Array and 53 pointings for the 7 m Array, with spacings of 13$\farcs$6 and 23$\farcs$2, respectively. Observations were conducted in May 2015 with the 12-m Array and in August 2014 with the 7-m Array. The observation parameters, summarized in Tables \ref{tab:obs}
 and \ref{tab:obs2}, are nearly identical to those of the Oph B2 observations reported by \citet{kamazaki19}.

We used four basebands for the observations. Two of these were dedicated to 1.3 mm dust continuum observations in dual-polarization mode, each with a correlator bandwidth of 2000 MHz, divided into 128 channels. After removing data near the baseband edges, the total effective bandwidth was approximately 7.4 GHz. The remaining two basebands were used for molecular line observations ($^{12}$CO J=2--1, $^{13}$CO J=2--1, C$^{18}$O J=2--1) in dual-polarization mode, with a correlator bandwidth of 117.2 MHz and 3840 channels.

The uv ranges sampled by the 12-m Array and 7-m Array data were 12.5 -- 348 $k \lambda $ and 8.1 -- 48 $k \lambda $, respectively. The minimum uv distance of the combined data corresponds to a maximum spatial scale of 25\arcsec. The calibration was carried out through observations of quasi-stellar and solar system objects. 
The details of the calibrators are listed in Table \ref{tab:obs}.
We used the Butler-JPL-Horizons 2012 model with the absolute flux scale of Titan and Mars.

We used the Common Astronomy Software Applications package, CASA pipeline version 4.2.2 and version 4.5.3, for calibration and imaging, respectively \fn{\citep{casa22}}. 
The calibration was carried out ourselves using scripts prepared by the ALMA observatory. 
As with the Oph B2 data analysis, we slightly modified the scripts for the 7-m Array data, where the shadowing criterion for the bandpass data was reduced from 7 m to 6 m to recover some data previously flagged by the original scripts. 
The CASA task {\it statwt} was used to determine  weights of the visibility data and  
the 12-m Array and 7-m Array data were combined using the CASA task {\it concat}. 
The multiscale clean \citep{cornwell08} option of {\it tclean} for the mosaic imaging of our multifield observations was applied. 
We adopted natural weighting. 
The final images have a beam size of approximately 1$\farcs$4$\times$0$\farcs$9. Additionally, we created images smoothed to a 2$\arcsec \times 2\arcsec$ resolution using the {\it imsmooth} task in CASA.
A part of the Oph A data were presented in \citet{kawabe18}, \citet{yamagishi19}, and \citet{hara21}.

\subsection{JWST Observations}

\fn{The Ophiuchi A region was observed with the Near Infrared Camera (NIRCam) onboard the James Webb Space Telescope (JWST) as part of the mission’s first anniversary imaging campaign (Program ID: 2739, PI: K. Pontoppidan). 
Observations were carried out over three epochs—March 7, April 5, and April 6, 2023—covering an area of approximately 49 square arcmin.}

\fn{Six filters were employed during the observations: F115W, F187N, F200W, F335M, F444W, and F470N
\footnote{\url{https://webbtelescope.org/contents/news-releases/2023/news-2023-128}}. 
The filter names indicate the central wavelengths of the observations: 1.15, 1.87, 2.00, 3.35, 4.00, and 4.70 $\mu$m, respectively. The final character in each filter name, W or N, denotes the filter type: wide-band or narrow-band, respectively.
These were selected to trace key emission and scattering features including the hydrogen Paschen-$\alpha$ line (F187N), the 3.3$\mu$m PAH band (F335M), scattered starlight (F115W and F200W), the H$_2$ S(9) rotational line (F470N), and the CO fundamental rovibrational band (F470N). 
The observational setup and data processing were largely based on the methodology described in \citet{pontoppidan22}, ensuring consistency with other JWST imaging campaigns.
}

We retrieved the \fn{stage 3} imaging data \fn{products} from the Mikulski Archive for Space Telescopes (MAST), which was publicly released on July 12, 2023. 
\fn{The images are visually inspected to mask artifacts and then coadded using the \texttt{reproject} package\footnote{\url{https://reproject.readthedocs.io/en/stable}}.}
The astrometry was aligned to the GAIA Data Release catalog \citep{gaia16,gaia23} for improved positional accuracy.

For this study, we specifically used the F470N data, chosen for its sensitivity to the H$_2$ 
$v=0-0$  S(9) line at 4.69 $\mu$m. This narrow-band filter is dominated by the H$_2$ emission line \citep{ray23}, which is excited in shock-heated regions. As such, it is well-suited for detailed investigations of stellar feedback phenomena, including protostellar jets.
\fn{The image resolution at 4.7 $\mu$m corresponds to 0.14\arcsec (or 20 au at the distance of Oph A).}

\begin{figure*}[htbp]
    \centering
        \includegraphics[width= 0.9 \textwidth]
        {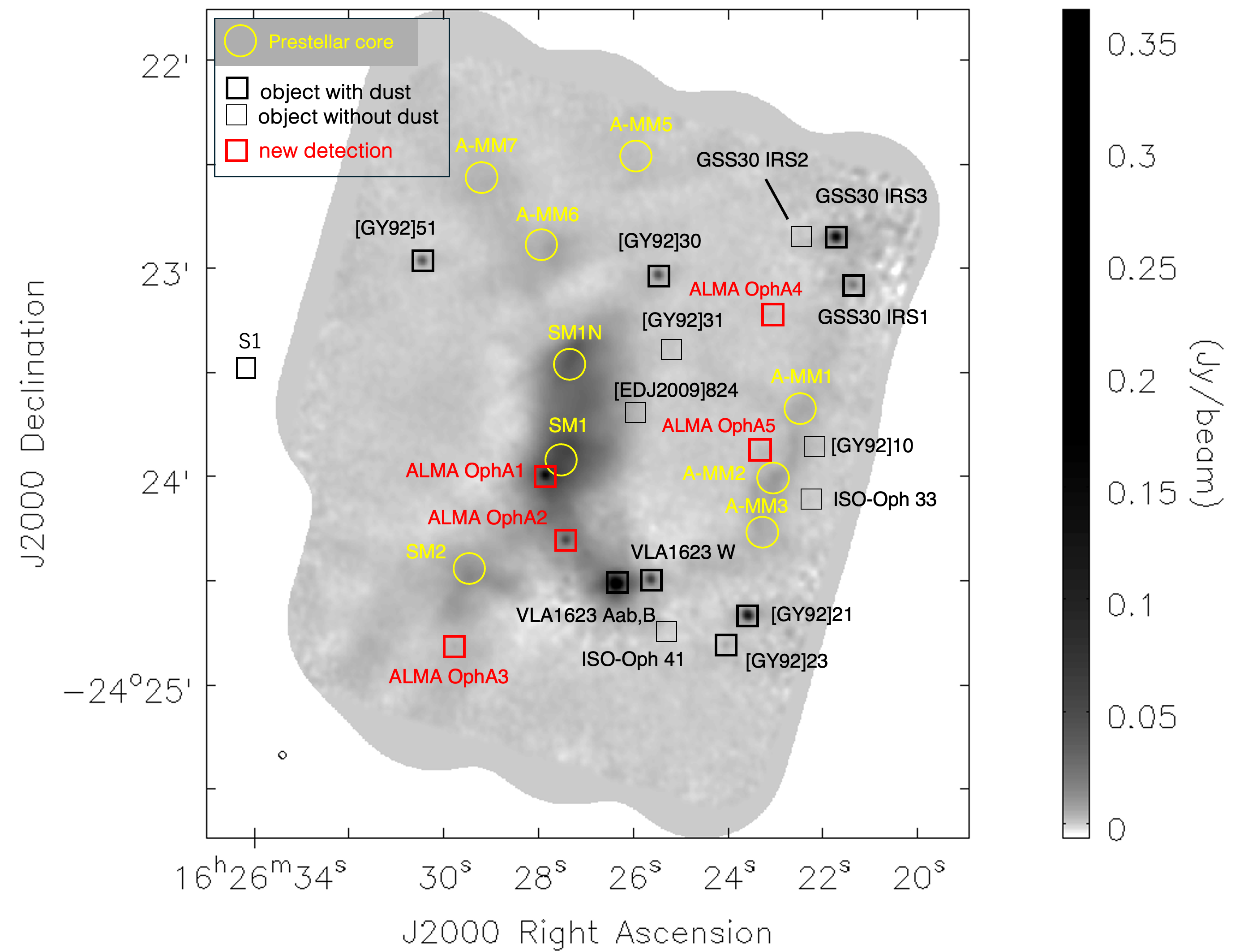}
\caption{1.3 mm dust continuum image of Oph A, created by combining the 7-m and 12-m Array data.  
The grayscale bar represents intensity in Jy~beam$^{-1}$.  
The image has been smoothed to an angular resolution of $2\arcsec \times 2\arcsec$.  
The synthesized beam is shown as an open circle in the bottom-left corner.  
The positions of prestellar cores identified by \citet{motte98} are marked with yellow circles.  
\fn{Known young stellar objects (YSOs) and newly discovered sources identified in our ALMA data are indicated by black and red squares, respectively.  
Sources associated with 1.3 mm continuum emission are shown with bold symbols.  
Newly discovered substellar objects without infrared emission are not marked here, as they are difficult to discern in the image.  
See Section~\ref{subsec:condensation} for detailed continuum images of these sources.}  
For reference, the Herbig Be star S1 is also marked with a square.}

   \label{fig:continuum}
\end{figure*}

\begin{figure*}[htbp]
    \centering
        \includegraphics[width= 0.9 \textwidth]
        {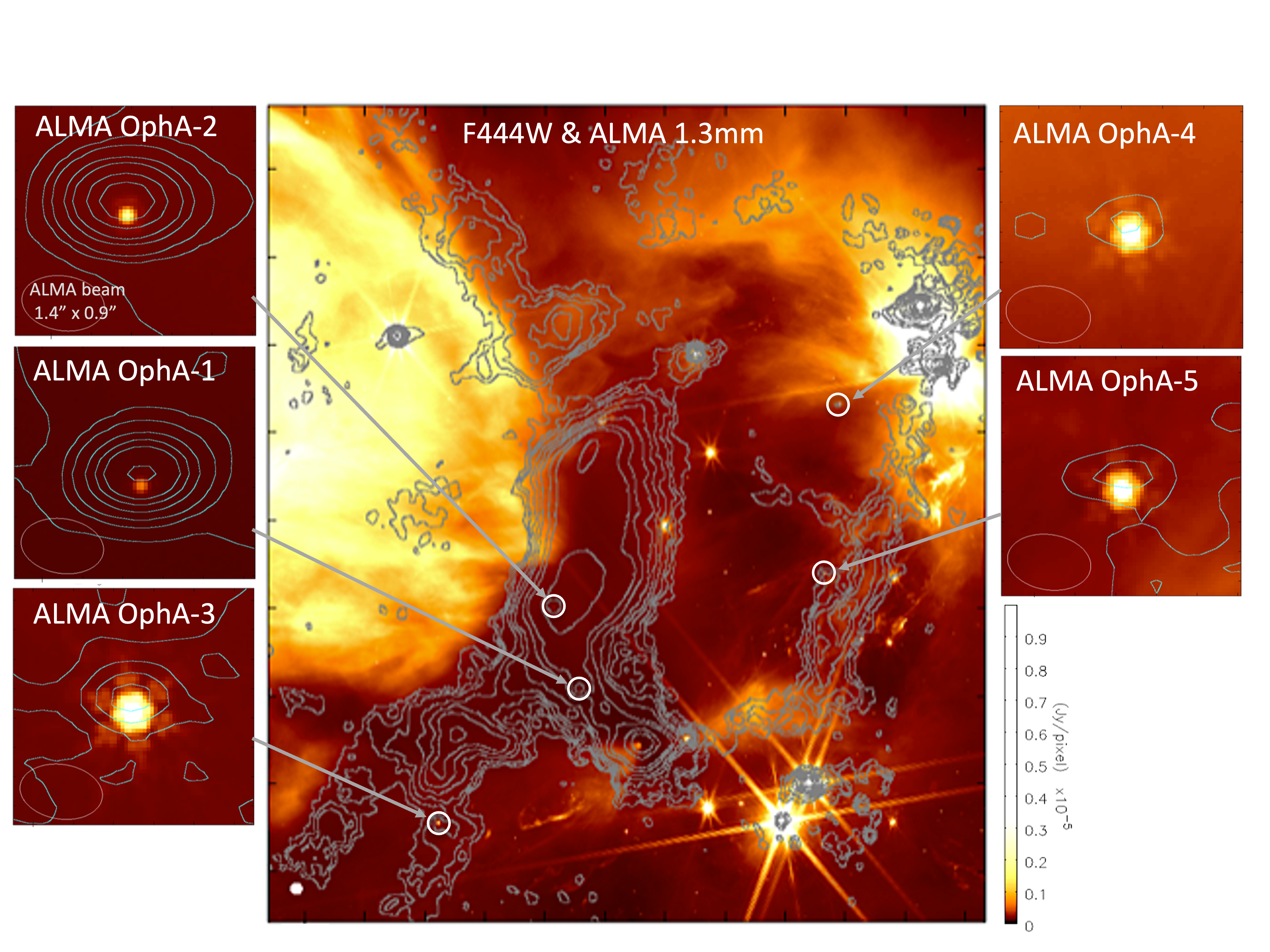}
\caption{\fn{Positions and close-up views of the newly detected infrared sources. The images present JWST F444W infrared emission (color scale) overlaid with 1.3 mm dust continuum contours.  
The central large panel shows a wide-field view with contours of the smoothed continuum image. The smaller panels display $5\arcsec \times 5\arcsec$ zoomed-in views, each overlaid with contours from the original image.  
The synthesized beam is indicated in the bottom-left corner of each panel.  
Contour levels are drawn at 3, 5, 7, 9, 11, 15, 20, 30, $\cdots$, 100, 200, and 300$\sigma$, where $1\sigma$ corresponds to 0.25 mJy beam$^{-1}$ for the original image and 0.445 mJy beam$^{-1}$ for the smoothed image (see Table~\ref{tab:obs2}).}}
    \label{fig:alma-bd-pmo}
\end{figure*}

\begin{figure*}[htbp]
    \centering
\includegraphics[width=\textwidth]{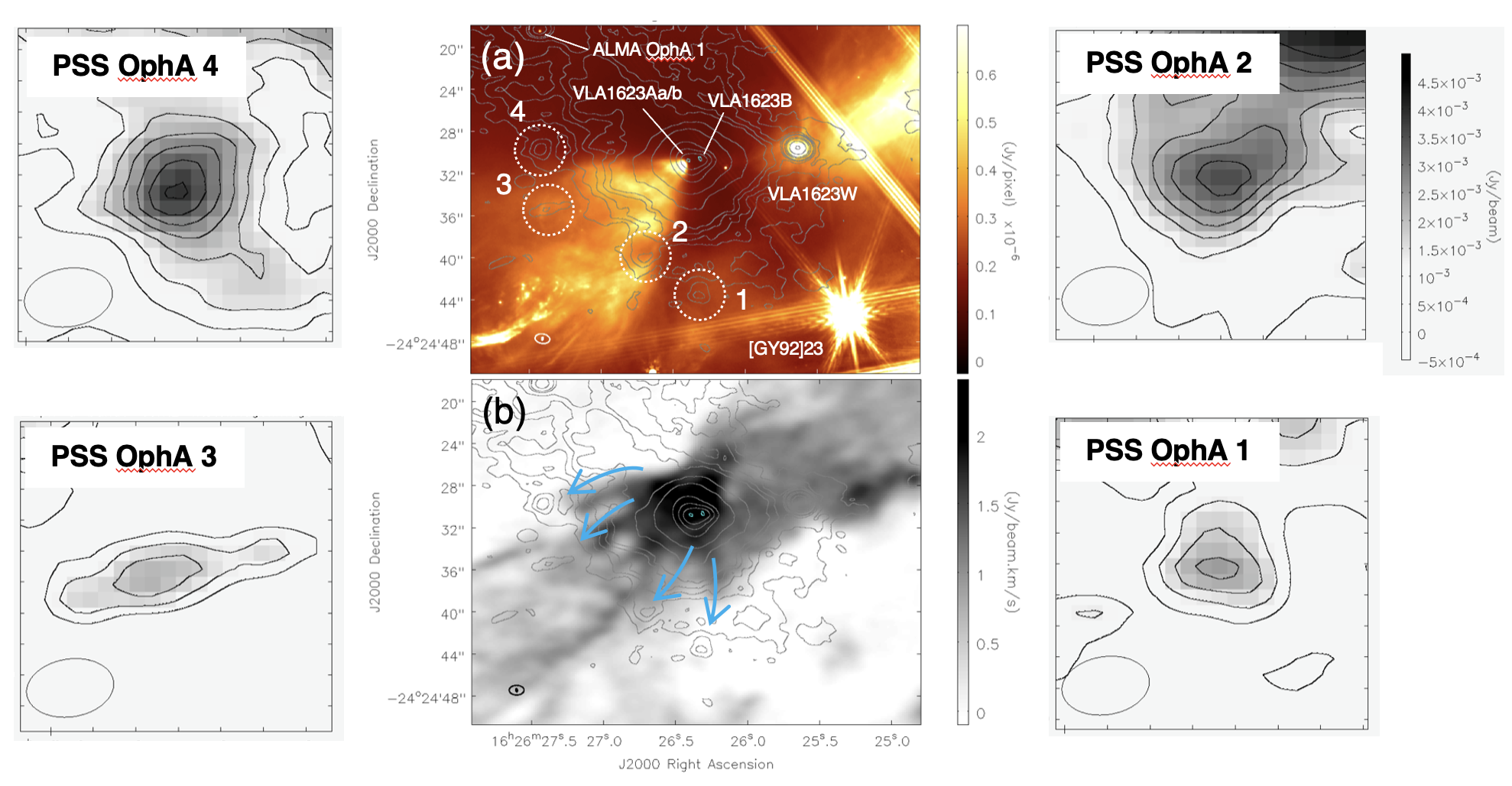}
\caption{\fn{Positions and close-up views of newly identified cores without point-like infrared emission, i.e., pre-substellar (PSS) core candidates.  
(a) JWST near-infrared image overlaid with 1.3 mm dust continuum contours;  
(b) ALMA 1.3 mm dust continuum contours overlaid on the $^{12}$CO ($J=2-1$) integrated intensity map around VLA 1623.  
The smaller panels surrounding (a) and (b) display zoomed-in views of $5\arcsec \times 5\arcsec$. 
PSS core candidates are marked with white dashed circles in panel (a).  
The positions of VLA 1623Aa/b and VLA 1623B are indicated in cyan, based on VLA 41 GHz continuum data \citep{kawabe18}.  
Dust continuum contours are shown in gray at levels of 4, 6, 8, $\cdots$ $\sigma$.  
In panel (b), the $^{12}$CO emission is integrated over velocities from 1.0 to 1.5 km~s$^{-1}$.  
Finger-like CO structures, highlighted by cyan arrows, extend outward from VLA 1623, seemingly pointing toward the PSS cores.  
In each small panel, dust continuum contours are drawn at 3, 5, 7, $\cdots$ $\sigma$.  
The synthesized beam is shown in the bottom-left corner of each panel.}
}

    \label{fig:condensation}
\end{figure*}

\section{Results}
\label{sec:result}


\begin{longtable*}{lccccccl} 
  \caption{Young Stellar Objects in the Observed Area.} \\
 \hline \hline 
Name$^a$ & Class & R. A.$^b$ & Decl.$^b$ & Dust$^c$ & Outflows$^c$ & Jets$^c$ &  Other Name$^d$  \\ 
 \hline 
\endfirsthead

\caption*{Table \thetable{} -- continued from previous page} \\
 \hline \hline 
Name$^a$ & Class & R. A.$^b$ & Decl.$^b$ & Dust$^c$ & Outflows$^c$ & Jets$^c$ &  Other Name$^d$  \\ 
 \hline 
\endhead

\hline \multicolumn{8}{r}{Continued on next page} \\
\endfoot

\hline
\endlastfoot

\multicolumn{2}{l}{Inside the ALMA mosaic area$^e$} & 16$^h$  & --24$^\circ$ & 1.3 mm & CO & IR \\ \hline
GSS30-IRS1 & I & 26$^m$21$^s$.361  & 23$\arcmin$:4\farcs78 & Y& BR & Y &  ISO-Oph29, [GY92]6, [EDJ2009]811, \\  
 & &   & & & & & \hspace{0.3cm} Elias 2-21, SSTc2d J162621.3-242304, \\  
  & &   & & & & &  \hspace{0.3cm} 2MASS J16262138-2423040 \\  
GSS30-IRS3  & I & 26:21.723 & 22:50.88 & Y & BR & Y & LFAM-1, ISO-Oph31,  \\ 
  &  &  & & & & & \hspace{0.3cm} SSTc2d J162621.7-242250,  \\ 
    &  &  & & & & & \hspace{0.3cm} 
2MASS J16262177-2422513  \\ 
  GSS30-IRS2 & II & 26:22.38  & 22:29 & N & N & Y & $[$GY92$]$12, LFAM-2, VGSS12,  \\  
& &   &  & & & & \hspace{0.3cm} SSTc2d J162622.4-242253, \\  
& &   &  & & & & \hspace{0.3cm} 2MASS J16262238-2422529 \\  
ISO-Oph 33 & F &26:22.26912 & 24:07.0632  & N & N & Y & [GY92]11, $[$EDJ2009$]$814,\\  
 &  & &   & & & & \hspace{0.3cm} SSTc2d J162622.2-242407, \\  
     &  &  & & & & & \hspace{0.3cm} 
2MASS J16262226-2424070 \\ 
 $[$GY92$]$10 & ? & 26:22.18584 & 23:52.3572  & N  & N & 
 N & 
2MASS J16262218-2423523 \\
$[$GY92$]$21 & F & 26:23.581 & 24:39.92  & Y & B & Y & LFAM-3, ISO-Oph 37, [EDJ2009]817,  \\  
&  & & &  & && \hspace{0.3cm}  SSTc2d J162623.5-242439, \\  
&  & & &  & && \hspace{0.3cm} 
2MASS J16262357-2424394 \\  
$[$GY92$]$23 & II & 26:24.045 & 24:48.44 & Y & B & Y &  ISO-Oph39, GSS 32, [EDJ2009] 820,  \\
 &  & &  &  &  & &  \hspace{0.3cm} Elias 2-23, SSTc2d J162622.4-242253, \\
 &  & &  &  &  & &  \hspace{0.3cm} 2MASS J16262404-2424480 \\
$[$GY92$]$31  & II & 26:25.23384 & 23:23.9208  & N & N & & 2MASS J16262523-2423239\\   
ISO-Oph 41 & II & 26:25.28280 & 24:45.0072 & N & N & N & [GY92]29, SSTc2d J162625.3-242445,   \\ 
 &  & &  &  &  & &  \hspace{0.3cm} 2MASS J16262528-2424450 \\
$[$GY92$]$30 & I & 26:25.474  & 23:1.81 & Y & BR & Y &	[EDJ2009]822, SSTc2d J162625.4-242301, \\  
 &  & &  &  &  & &  \hspace{0.3cm} 
2MASS J16262548-2423015 \\
VLA 1623W & I & 26:25.630 & 24:29.55 & Y & B? & Y? & 	[EDJ2009] 823, SSTc2d J162625.6-242429  \\  
$[$EDJ2009$]$824 & F & 26:25.99 & 23:40.5 & N & 
N & N &  SSTc2d J162625.9-242340 \\  
VLA 1623B$^e$ & 0 & 	26:26.308 & 24:30.601 & Y & ? & N? & LFAM5, [EDJ2009]825,  \\  
 &  & 	 &  &  & & & \hspace{0.3cm} SSTc2d J162626.4-242430  \\  
VLA 1623Aa$^e$ & 0 & 26:26.399 & 24:30.721 & Y & BR & Y & LFAM5, [EDJ2009]825,  \\  
VLA 1623Ab$^e$ & & 26:26.386  &  24:30.765 &  & & & \hspace{0.3cm} SSTc2d J162626.4-242430  \\  
\hline
ALMA OphA1 & 0/I? & 26:27.424 & 24:18.27 & Y & R? & N & Source X$^i$ (see also Figure \ref{fig:alma-bd-pmo}) \\  
ALMA OphA2  & 0/I & 26:27.854 & 23:59.54 & Y & B & N & SM1, SM1-A (see also Figure \ref{fig:alma-bd-pmo}) \\
{\bf \footnotesize ALMA OphA3} & I? &26:19.777 & 24:48.684 & Y & B? & N & substellar object (see also Figure \ref{fig:alma-bd-pmo})\\  
{\bf \footnotesize ALMA OphA4} & I? & 26:23.09 & 23:13.3764 & Y & N & N & substellar object (see also Figure \ref{fig:alma-bd-pmo})\\  
{\bf \footnotesize ALMA OphA5} & I? & 26:23.3367 & 23:51.7906 & Y & N & N & substellar object (see also Figure \ref{fig:alma-bd-pmo})\\  
$[$GY92$]$51 & F/II & 26:30.452  & 22:57.73 & Y & N & N & LFAM9, VSSG27, ISO-Oph46, \\  
 &  &   &  & & & & \hspace{0.3cm} SSTc2d J162630.4-242257, \\  
  &  &   &  & & & & \hspace{0.3cm} 2MASS J16263046-2422571 \\  
\hline
\multicolumn{3}{l}{Outside the ALMA mosaic area} \\ \hline
S1  & Herbig Be & 26:34.175044 & 23:28.33187 & & & & [GY92]70, GSS35, ISO-Oph 48, Elias 2-25, \\
& /B4V &  &  & && & \hspace{0.3cm} SSTc2d J162634.2-242328,  \\
  &  &  &  & & &  & \hspace{0.3cm} 2MASS J16263416-2423282  \\
GSS31  & II & 26:23.40 & 21:00.8 & & & Y & ISO-Oph 36,  [GY92]20, [EDJ2009]816,  \\
  &  &  &  & & &  & \hspace{0.3cm} Elias 2-22, SSTc2d J162623.40-242100.8,  \\
    &  &  &  & & &  & \hspace{0.3cm} 2MASS J16262335-2420597  \\
ISO-Oph 21 & I & 26:17.23 & 23:45.1 & & & Y & [EDJ2009] 803, \\
 &  &  & & & &  &  SSTc2d J162617.23-242345.1, \\
   &  &  &  & & &  & \hspace{0.3cm} 2MASS J16261722-2423453  \\
ISO-Oph 26 & II/F & 26:18.98 & 24:14.2 & & & Y & 	[EDJ2009] 807,  \\
 &  &  &  & & & & \hspace{0.3cm}	 SSTc2d J162618.98-242414.2, \\
   &  &  &  & & &  & \hspace{0.3cm} 2MASS J16261898-2424142   \\
GSS29  & II & 26:16.84 & 22:22.9 & & & Y & ISO-Oph 19, 	[EDJ2009] 801,  \\
  &  &  &  & & &  & \hspace{0.3cm} SSTc2d J162616.84-242222.9,  \\
    &  &  &  & & &  & \hspace{0.3cm} 
2MASS J16261684-2422231  \\
$[$GY92$]$20  & II & 26:23.36845 & 20:59.5794  & & & Y & ISO-Oph 36, [EDJ2009] 816,  \\
  &  &  &  & & &  & \hspace{0.3cm} SSTc2d J162623.4-242101,  \\
    &  &  &  & & &  & \hspace{0.3cm} 
2MASS J16261684-2422231  \\
\hline
 \hline 
\end{longtable*}
\tablecomments{
\fn{$^a$The names with ALMA OphA represent the objects detected by our 1.3 mm dust continuum map and ALMA OphA 1 and 2 are reported in \citet{kawabe18}. The object names shown in bold are the objects newly detected in this paper.}
$^b$The positions of the YSOs are determined either by optical or near-infrared observations and 
basically adopted from the SIMBAD astronomy database (http://simbad.u-strasbg.fr).
The equinox of the coordinate system is J2000.
The position of ALMA OphA 2, also referred as SM1 and SM1-A in previous studies, is adopted from \citet{kawabe18} who classified it as a protostellar core.
\fn{$^c$The 5th, 6th, and 7th columns represent the detections of the ALMA 1.3 mm dust continuum emission, CO outflow, and IR jet/knots components, respectively. Y and N indicate the detection and non-detection, respectively. In the 6th column, B and R represent CO blueshifted and redshifted component detections, respectively.}
$^d$The references for other names are as follows: GSS: \citet{grasdalen73}, GY92: \citet{greene92}, ISO-Oph: \citet{bontemps01}, 
Elias: \citet{elias78}, LFAM: \citet{leous91}, VSSG: \citet{vrba75}, EDJ2009: \citet{evans09}. 
All the known YSOs in the ALMA mosaic area are listed in the table.
For CO outflows, see also \citet{kamazaki01,nakamura11,white15}.
$^e$ Based on the VLA 41 GHz data in \citet{kawabe18}.
}
\label{tab:objects}

\begin{table*}[htbp]
  \centering
  \caption{Prestellar Cores in the Observed Area.}
  \begin{tabular}{llll}
    \hline \hline
    Name & R. A. & Decl. & Comment \\
    \hline
    $[$MAN98$]$ A-MM1  & 26:22.44   & 23:40.2   & \\
    $[$MAN98$]$ A-MM2      & 26:23.04   & 24:00.1   & \\
    $[$MAN98$]$ A-MM3      & 26:23.25   & 24:16.1   & \\
    $[$MAN98$]$ A-MM5      & 26:25.91   & 22:26.9   & \\
    $[$MAN98$]$ A-MM6      & 26:27.93   & 22:52.8   & \\
    $[$MAN98$]$ A-MM7      & 26:29.22   & 22:33.7   & \\
    $[$MAN98$]$ A-SM1N     & 26:27.34   & 23:27.9   & \\
    $[$MAN98$]$ A-SM1      & 26:27.55   & 23:55.8   & SM1-A, ALMA OphA 2, contain a Class 0 object \\
    $[$MAN98$]$ A-SM2      & 26:29.46   & 24:26.7   & \\
    \hline
    \multicolumn{4}{l}{\footnotesize Pre-substellar cores around VLA 1623 (see also Figure~\ref{fig:condensation})} \\
    \hline
    \textbf{PSS OphA1} & 26:26.3154 & 24:43.3787 &  \\
    \textbf{PSS OphA2} & 26:26.6858 & 24:39.7420 &  \\
    \textbf{PSS OphA3} & 26:27.365  & 24:35.332  & \\
    \textbf{PSS OphA4} & 26:27.4148 & 24:29.6574 & \\
    \hline \hline
  \end{tabular}
\label{tab:prestellar cores}
  \tablecomments{
The objects with [MAN98] are the cores identified from the 1.2 mm continuum observations done by \citet{motte98}. 
}
\end{table*}

\begin{figure*}[htbp]
    \centering
\includegraphics[width= 0.8 \textwidth]{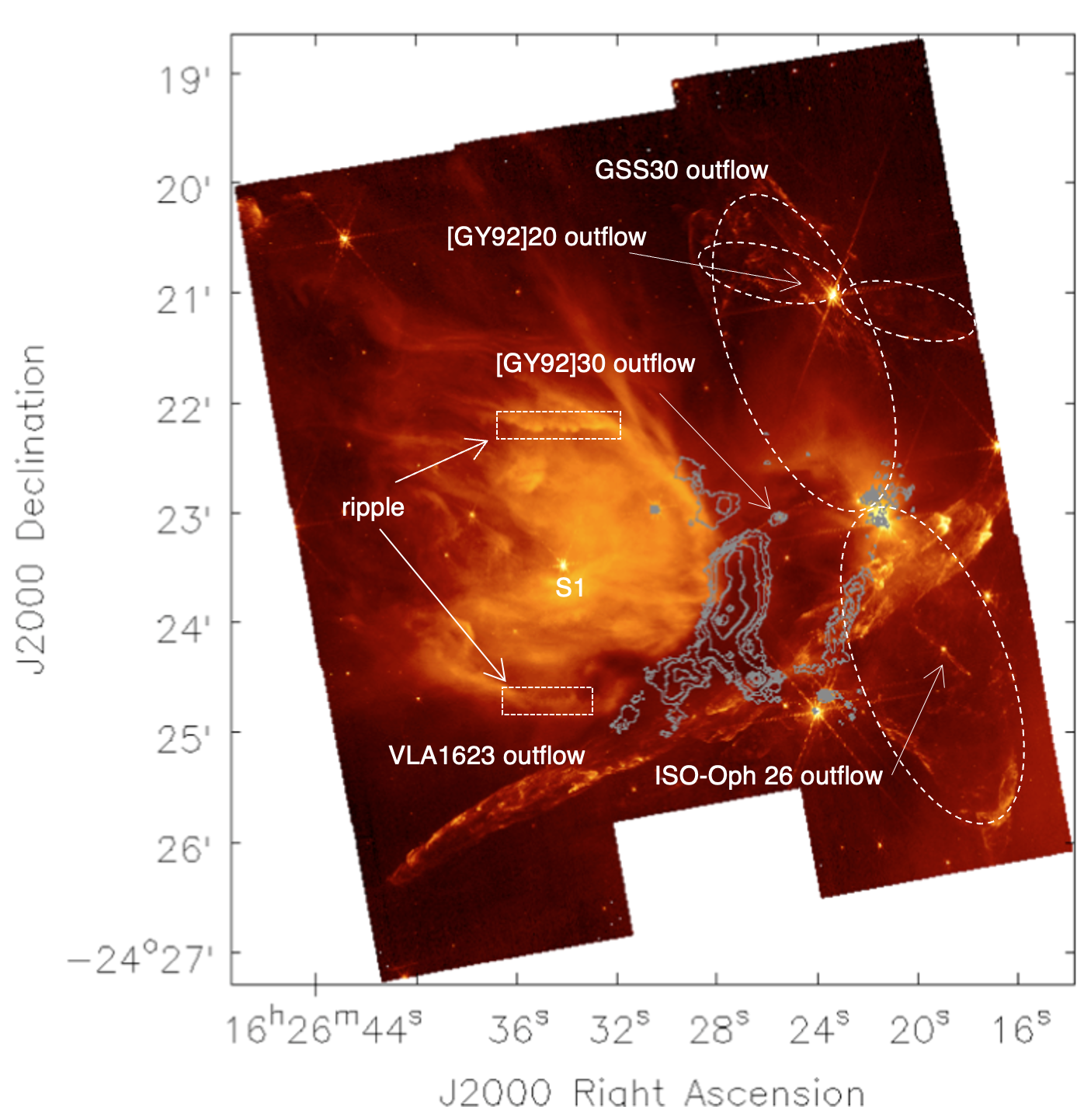}
    \caption{\fn{Wide-field} JWST NIRCam F470N image, overlaid with the contours of the smoothed 1.3 mm continuum emission. The contours are drawn at 6, 12, 24, 60, 180, and 240 $\sigma$ with $1 \sigma = $ 0.45 mJy/beam. The outflows of VLA1623, [GY92]30, GSS 30, ISO-Oph 26, and \fn{[GY92]20} are indicated. \fn{The close-up view of the [GY92]20 outflow is presented in Section \ref{subsubsec:outflows}.} The ripple structures possibly created by the Kelvin-Helmholtz instability are enclosed in dotted squares.}
    \label{fig:jwst}
\end{figure*}

\subsection{1.3 mm dust continuum image and YSOs/prestellar cores}

Figure \ref{fig:continuum} presents the combined ALMA 7 m Array and 12 m Array 1.3 mm continuum image of Oph A. 
The prominent elongated structure is the Oph A ridge, which is the densest subclump in $\rho$ Oph. The known young stellar objects (YSOs) and prestellar cores in this area are listed in \fn{Tables \ref{tab:objects} and \ref{tab:prestellar cores}, respectively}. \fn{Additionally, we have identified 7 faint objects in this study (ALMA OphA 3/4/5 and PSS OphA 1/2/3/4). We indicated ALMA OphA 3/4/5 and 2 low-mass objects (ALMA OphA 1/2) detected from our ALMA image in Figure \ref{fig:continuum} in red.
The comparison with the JWST F444W image is presented in Figure \ref{fig:alma-bd-pmo}.
We also detected 4 faint cores lack associated compact infrared emission around VLA 1623 (PSS OphA 1/2/3/4), whose enlarged views are presented in Figure \ref{fig:condensation}. }

\fn{The point-source sensitivity at $4.4 \ \mu m$ data is derived to be $m_{5-10 \sigma} \sim $ 24 mag for a $5-10 \sigma$ detection. 
An object with an intrinsic (unextincted) magnitude $m_0$ at $4.4 \ \mu m$ appears fainter due to extinction, and its observed magnitude $m_{\rm obs}$ in a region with visual extinction $A_V$ is given by
$m_{\rm obs} = m_0+0.054 A_V$ at $4.4\mu m$. 
Accordingly, the maximum visual extinction $A_V$ at which an object can be detected is
\begin{equation}
A_V \lesssim \frac{m_{5-10 \sigma}-m_{\rm obs}}{0.054}
\end{equation}
For example, a $10 \ M_{\rm Jup}$ object at an age of 1 Myr has a observed magnitude ($m_{\rm obs}$) of about 13.7 mag at $4.4 \ \mu m$ from Figure 3 of \citet{defurio25} correcting the distance difference (400 pc $\rightarrow $ 138 pc). 
Such an object would be detectable in regions with $A_V \lesssim 191$.
Similarly, objects with masses greater than $10 M_{\rm Jup}$ remain detectable in areas with $A_V \sim 10$ up to an age of a few Myr.
Even for objects with lower masses, e.g., a 1 $M_{\rm Jup}$ object with 1 Myr, the detection limit is calculated to be $A_V \lesssim (24-17.7)/0.054 \sim 117$.
Though less massive objects are more difficult to be detected, the Jupiter mass objects are detectable in low $A_V$ area, if exists.}

\fn{Given the significant extinction in Oph A (typically $A_V = 10$–200), even the high sensitivity of JWST may not be sufficient to detect faint infrared emission from deeply embedded sources.
Therefore, we classify the four undetected cores as pre-stellar or pre-substellar (PSS) candidates, likely located in regions with $A_V \sim 10$–200.}

\fn{Further analysis of these small cores is presented in Section~\ref{subsec:condensation}.
The nine newly discovered objects are listed in bold in \fn{Tables~\ref{tab:objects} and \ref{tab:prestellar cores}}
\fn{See Section~\ref{subsec:condensation} for the details of the identification.}
For reference, the location of the Herbig Be star S1, which lies outside the ALMA field of view, is indicated in Figure~\ref{fig:continuum}.}

All Class 0 and Class I objects in the area are associated with compact dust continuum emission.
On the other hand, among the Flat Spectrum sources and Class II objects, some have compact dust continuum emission, while others do not.
Most protostars are associated with point-like JWST infrared emission; however, a few objects, such as VLA 1623Aab (which only shows scattered light) and VLA 1623B, are not detected in the JWST infrared data. All the Flat Spectrum and Class II objects have point-like JWST emission. 
The positions of prestellar cores identified by \citet{motte98} are shown as yellow circles onto the combined 7 m and 12 m Array image. Most prestellar cores have either very weak or undetectable emission, likely due to interferometric spatial filtering. 
Comparison with the 1.2 mm continuum image \citep[see Figure 1 of][]{motte98} 
highlights the significant challenge of identifying prestellar cores, which have extended density structures, using only ALMA interferometric data.
This missing flux problem from spatial filtering is common in ALMA data, and we confirmed it in the continuum image of Oph B2 (the second densest subclump in $\rho$ Oph) as well \citep{kamazaki19}.



\subsection{JWST near-infrared image}

Figure \ref{fig:jwst} presents the wide-field JWST NIRCam F470N image of the $\rho$ Oph A region. 
For comparison, the ALMA dust continuum emission presented in Figure \ref{fig:continuum} is shown as contours.
Several point-like objects are found in the JWST image, either young stars or protostars.
In this F470N band ($4.7 \mu$m), the molecular hydrogen emission line $v=0-0$ S(9) at $4.69 \mu$m is a powerful tracer of shock-excitation and thus protostellar jets and knots. 
Molecular hydrogen emission also becomes strong within photon-dominated regions (photo-dissociated regions, PDRs), the transition zones between dense, cold molecular gas and the tenuous, warm, ionized gas \citep{hollenbach99,kristensen23}. 
In the JWST image in Figure \ref{fig:jwst}, most of the strong emission comes from either shocked areas or PDRs.

The brightest nebula in the eastern part originates from S1, the most luminous young star in the area, which is classified as a Herbig Be star. S1 is a compact binary system with a separation of a few au. The primary star has a mass of about 4 M$_\odot$, while the companion has a mass of about 0.8 M$_\odot$ \citep{jazmin24}. This young intermediate-mass pre-main-sequence star excites an H{\sc ii} region and illuminates a large, egg-shaped PDR \citep{mookerjea21}. The PDR appears to be confined to the west and southwest by the dense molecular Oph A ridge which is seen in the 1.3 mm dust continuum emission, while it expands more freely into the diffuse, low-density cloud to the northeast.

\fn{Within two regions outlined by the dotted white boxes, ripple-like structures are apparent. These features closely resemble the patterns attributed to the Kelvin-Helmholtz instability discovered at the surface of the H{\sc ii} bubble in Orion A \citep{berne10,berne12}.}

In this infrared image, several Herbig-Haro (H-H) objects or knots, formed by shocks from protostellar jets, are observed.
The YSOs associated with these infrared H-H objects are indicated in Table \ref{tab:objects}.
In Figure \ref{fig:jwst}, we indicate several prominent protostellar jets. 
The most prominent H-H object, seen in the collimated jet-like emission, is
the one associated with the Class 0 object, VLA 1623. 
Previous infrared observations revealed a chain of infrared knots associated with this protostellar jet \citep[see e.g.,][]{khanzadyan04, zhang09}. Such chain knots are seen in this image.


\begin{table*}[htbp]
 \centering
  \caption{Observed properties of small and compact dust cores near VLA 1623}
  \begin{tabular}{lcccccc}
 \hline \hline 
ID &  R. A. & Dec. & Peak flux & Integrated flux & \fn{Infrared} & FWHM major $\times$ minor axes \\ 
& (J2000)  & (J2000) & (mJy/beam) & (mJy) &   & (\arcsec \ $\times$ \arcsec)  \\ 
 \hline 
1 &  16:26:26.3154  & --24:24:43.3787 & 2.36 $\pm$ 0.33 &  6.1 $\pm$ 1.4 & N & (1.64 $\pm$ 0.4) $\times$ (1.21 $\pm$ 0.66) \\  
2 &  16:26:26.6858  & --24:24:39.7420 & 3.2 $\pm$ 0.3 & 21.9 $\pm$ 2.3 &N  & (3.06 $\pm$ 0.38) $\times$ (2.48 $\pm$ 0.31)  \\ 
3  &  16:26:27.365 & --24:24:35.332 & 2.19 $\pm$ 0.36 &  5.2 $\pm$ 1.3& N& (3.58 $\pm$ 1.14) $\times$ (0.87 $\pm$ 0.11) \\ 
4 & 16:26:27.4148 & --24:24:29.6574 & 3.24 $\pm$ 0.38 & 17.4 $\pm$ 2.4& N  & (2.71 $\pm$ 0.04) $\times$ (2.0 $\pm$ 0.042)\\  \hline
ALMA OphA 3 & 16:26:19.777 & --24:24:48.684 & 2.38 $\pm$ 0.17 & 4.96 $\pm$ 0.51 & Y  & (1.33 $\pm$ 0.28) $\times$ (1.01 $\pm$ 0.30)   \\
ALMA OphA 4 & 16:26:23.09 & --24:23:13.3764  & 1.43 $\pm$ 0.15 & 1.39 $\pm$ 0.27 & Y  & point source   \\
ALMA OphA 5 & 16:26:23.3367 & --24:23:51.7906 & 1.46 $\pm$ 0.10 & 1.90 $\pm$ 0.39 & Y  & (1.28 $\pm$ 0.57) $\times$ (0.16 $\pm$ 0.33)   \\
 \hline 
\end{tabular}
  \label{tab:condensation}
\tablecomments{
The major and minor axes are measured by a two-dimensional Gaussian fit using CASA.
These radii represent beam-deconvolved values except for core 3.
}
\end{table*}

A large bow-shaped H-H object extends in the north-south direction, as seen in the right-hand side of the panel. It is presumably driven by the outflow from  GSS 30 IRS1, judging from its bipolar morphology as indicated in Figure \ref{fig:jwst}.
\fn{In the figure, we indicated the bow structures with two identical dashed ellipses for reference.}  
GSS 30 IRS1 is a bright low-mass Class I binary system with a separation of 21 au \citep{chen07}.
From the image, a jet-like structure extending from IRS 1 is also seen (see Section \ref{subsubsec:outflows} (Figure \ref{fig:outflows}(a)) for more details.).
GSS 30 IRS 2 and IRS 3 are also protostars with associated molecular outflows, which are also aligned in the north-south direction (see Section \ref{subsubsec:outflows} for details), and might contribute to the structures created by protostellar jets.
The H-H object is extended about 6\arcmin, corresponding to 0.25 pc.
\citet{zhang09} detected a part of the southern nebula in the {\it Spitzer} IRAC data and identified it as extended green object (EGO) 09. 
\citet{khanzadyan04} also detected the same feature in the near-infrared (f10-03 in their Figures 24 and 25). They suggest it is driven by a source to the north, judging from its bow shape. This interpretation is also consistent with the near-infrared reflection nebulae revealed by polarimetry \citep{tamura91}, which traces the outflow cavity that extends to both north-east and south-west.

A narrow shaft of emission seen near the northern edge of the Oph A ridge is the protostellar jet from the Class I object, [GY92]30 \citep{kamazaki01,nakamura11,white15}.
A small jet-like structure near the southwest corner is a protostellar jet from the Class II or Flat Spectrum object, ISO-Oph 26, which is located outside our ALMA image (see Figure A.23 of \citet{zhang13} for its proper motion measurement).  
\fn{In addition, we identified a bipolar jet from the Class II object, [GY92]20, whose shocked components are overlapped on the plane-of sky by the GSS 30 bow structure.}
In Section \ref{subsubsec:outflows}, we compare the jet-like infrared emission near the YSOs with the high-velocity $^{12}$CO emission, to identify the driving sources.  
In Table \ref{tab:objects}, the results of these identifications are indicated in the 7th column.

\section{Comparison between ALMA and JWST data}
\label{sec:comparison}

\begin{table*}
 \centering
  \caption{Derived properties of small and compact dust cores near VLA 1623}
\begin{tabular}{cccccc}
 \hline \hline 
ID & $M_c$$^a$ & $R_c$ &  $\left<n_c\right>$$^a$  & $M_{\rm BE}$$^a$ & $\alpha _{BE}$$^a$ \\ 
& ($\times 10^{-3}$ M$_\odot$)  & (au)  & ($\times 10^8$ cm$^{-3}$) & ($M_\odot$)    & \\ 
 \hline 
1 &  11.8 -- 4.98 & 97.2 & 1.4 -- 3.4  & 16.2 -- 20.2 & 0.25 -- 1.9 \\  
2 &  42.4$-$17.9 & 190.1    & 1.1 -- 2.6 & 12.2 -- 39.4 & 0.45 -- 3.5 \\  
3 &  10.0$-$4.25 & 121.8    & 0.98 -- 2.3 & 7.8 -- 25.3 & 0.17 -- 1.3 \\  
4 &  33.7$-$14.2 & 160.6     & 2.3 -- 5.4 & 10.3 -- 33.3 & 0.43 -- 3.3\\ \hline 
\scriptsize ALMA OphA 3 & 3.53  & --$^c$ & &  & \\
\scriptsize ALMA OphA 4 &  0.99 & --$^c$ & &  & \\
\scriptsize ALMA OphA 5 & 1.67  & --$^c$ & &  & \\
 \hline 
\end{tabular}
\tablecomments{
$^a$ The minimum and maximum values were determined assuming the gas temperatures of 10 K and 18 K, respectively. \\ 
$^c$ALMA-OphA 3/4/5 are point sources.
}
  \label{tab:condensation2}
\end{table*}

\subsection{Discovery of Substellar Cores}

\fn{Here, we identified cores through visual inspection, with a focus on faint compact structures located near the edges of larger-scale emission features such as filaments and ridges. This approach was chosen because our ALMA continuum data, comprising both 7m and 12m array observations, are susceptible to spatial filtering effects, which can bias automated source-finding algorithms, especially in regions with complex and spatially varying background emission. 
To ensure objectivity and reproducibility, we adopted the following observational criteria for defining a core: (1) The source must exhibit a peak intensity exceeding 8$\sigma$, and
(2) It must show more than two closed contours when the dust continuum map is contoured starting at 3$\sigma$ with 1.5$\sigma$ intervals. We will conduct a more comprehensive and systematic core survey in a future study, incorporating additional single-dish data, which will help recover extended emission and improve source identification using automated methods.}

\fn{Through visual inspection of the 1.3 mm ALMA dust continuum image and JWST images, we identified seven compact faint cores with 
(ALMA OphA 3/4/5) and without point-like infrared emission (PSS OphA 1/2/3/4). These positions and some properties are summarized in \fn{Tables \ref{tab:objects} and \ref{tab:prestellar cores}}.
Particularly the cores without point-like infrared emission are found in the vicinity of VLA 1623 (except for ALMA OphA 4/5).
The enlarged views are presented in Figures \ref{fig:alma-bd-pmo} and \ref{fig:condensation} for the objects with and without point-like infrared emission, respectively.}

These cores exhibit significant signal-to-noise ratios in 1.3 mm dust continuum emission, exceeding 8--5 $\sigma$, indicating that they likely represent real column density structures. 
Figure \ref{fig:condensation}(a) highlights these compact cores with dotted circles.
\fn{We also indicated the enlargement of these objects in Figures \ref{fig:alma-bd-pmo}
(ALMA OphA 3/4/5 + ALMA OphA 1/2)
and \ref{fig:condensation} (PSS OphA 1/2/3/4).}
While additional cores with peak intensities exceeding 8$\sigma$ are present near VLA 1623, we have chosen to limit our analysis to those with relatively well-defined, isolated structures. 
A more complete identification of these sources will be addressed in a future study.
\fn{We note the presence of several faint, compact structures near the edge of the western filament, where prestellar cores A-MM1 through A-MM4 are located. Although these structures resemble the identified cores morphologically, their peak intensities fall slightly below the 8$\sigma$ threshold, and thus they are not included in the present analysis.}

Among them, three cores in the southern and western regions of Oph A coincide with sources in the JWST near-infrared image, suggesting them as potential YSO candidates (ALMA OphA 3/4/5). The other four cores lack associated compact infrared emission. However, given the significant extinction in Oph A, even the highly sensitive JWST data may fail to detect faint emission from deeply embedded sources.
For this study, we classify the faint continuum cores based on their infrared detectability: cores with 4.7 $\mu$m JWST emission are referred to as protostellar, proto-substellar, or young substellar objects, while those without infrared counterparts are considered prestellar or pre-substellar cores. The observational properties of these identified cores are summarized in Table \ref{tab:condensation}.
\fn{See \citet{andre12} for identification of a similar object in Oph B.}

\fn{The objects with point-like infrared emission have small offsets from the dust continuum peaks in almost the same directions. This is due to the effect of the proper motions for about 8 years.  Preliminary proper motion analysis with infrared and optical data indicates that these are associated with Oph group rather than background objects (Kawabe et al. 2025 in prep.).}

\subsubsection{Pre-substellar cores}
\label{subsec:condensation}

To estimate the masses of these faint 1.3 mm continuum cores, we use the following relation:
\begin{equation}
M=\frac{\tau _\nu}{1-\exp (-\tau _\nu)} \frac{S_\nu D^2}{\kappa _\nu B_\nu (T_d)}  \ ,
\end{equation}
where $S_\nu$ is the 1.3-mm flux density, $D$ is the distance from the Sun, $B_\nu$ is the Planck function, $\tau_\nu$ is the optical depth which is assumed to be low ($<1$), $T_d$ is the dust temperature assumed to be 20 K, the opacity $\kappa_{230 \rm GHz} = 0.00529 $ cm$^2$ g$^{-1}$ with $\beta=2$ \citep{hildebrand83}, and the gas-to-dust ratio is set to 100.
The physical quantities of the cores are listed in Table \ref{tab:condensation2}.
The 1.3 mm flux densities of these cores range from 3 mJy to 13 mJy, corresponding to estimated masses of 0.01--0.035 M$_\odot$.
We note that the estimated mass was calculated, assuming a temperature of 20 K. However, if the temperature varies between 10 K and 30 K, the mass estimate can change by a factor of 3.64 to 0.61.
No point-like infrared emission is observed at the positions of these sources, except for \fn{three sources} \fn{presented in Figure \ref{fig:alma-bd-pmo}}).

To assess the gravitational boundedness of the four observed cores without associated infrared emission, we calculate the Bonnor-Ebert mass ratio,
\begin{equation}
    \alpha_{\rm BE}=\frac{M_c}{M_{\rm BE}} \ ,
\end{equation}
which is the ratio of each source's mass ($M_c$) to the critical Bonnor-Ebert mass ($M_{\rm BE}$). This critical mass represents the maximum mass of a stable, pressure-confined isothermal sphere before it becomes gravitationally unstable and begins to collapse. It is given by
\begin{equation}
M_{\rm BE} = \frac{2.4R_c c_s ^2}{G}  
\fn{=1.44 \times 10^{-2} M_\odot \left(\frac{R_c}{10^2 {\rm au}}\right) \left(\frac{T}{15 \rm K}\right)^2
}
\ ,
\end{equation}
where $R_c$ is the radius of the core, $c_s$ is the isothermal sound speed ($c_s \simeq 0.23 $ kms$^{-1}$ at 15 K), and $G$ is the gravitational constant.

Our results show that all four cores without associated infrared emission have Bonnor-Ebert ratios close to unity 
($\alpha_{\rm BE} \approx 1$), indicating that they are likely  gravitationally bound on their own. \fn{The  properties of the cores resemble those of the pre-brown dwarf candidate, Oph B-11, identified by \citet{andre12}.}
Therefore, these substellar-mass cores may represent pre-substellar cores. On the other hand, ALMA-OphA 3 has point-like infrared emission and has already undergone gravitational contraction, evolving into a young substellar object. We discuss some details of this young substellar object in Section \ref{subsub:brown drawf}.

\begin{figure}[htbp]
    \centering
\includegraphics[width=0.5 \textwidth]{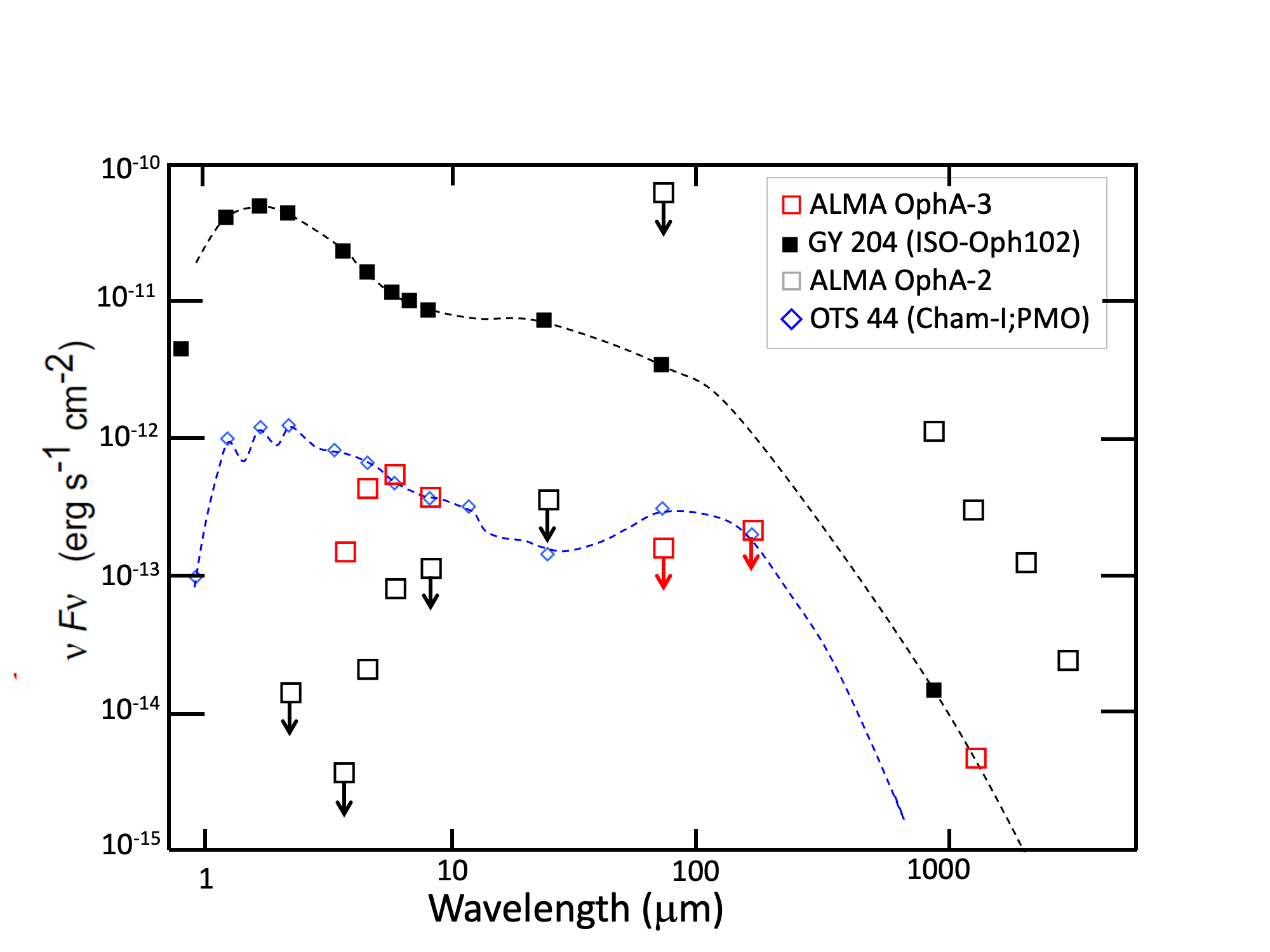}
\caption{
Spectral Energy Distributions (SEDs) of several low-mass objects.  
\fn{The red squares represent the SED of ALMA-OphA 3, which is compared with those of the following sources:  
the extremely young Class 0 object ALMA-OphA 2 (open squares; adapted from Figure 7 of \citet{kawabe18}),  
the young brown dwarf GY~204 (filled squares from \citet{Joergens13}, with its best-fit model shown as a black dashed line),  
and the free-floating planetary-mass object OTS~44 (blue diamonds from \citet{natta02}, with its best-fit model shown as a blue dashed line). 
The fluxes for ALMA-OphA 3 are derived from multiple instruments:  
{\it Spitzer} IRAC (3.6, 4.5, 5.8, and 8 $\mu$m; obtained from the NASA/IPAC Infrared Science Archive),  
{\it Herschel} PACS (70 $\mu$m; from the {\it Herschel} Science Archive), and ALMA (1.3 mm; this work).}}
    \label{fig:sed}
\end{figure}

\begin{figure*}[htbp]
    \centering
\includegraphics[width=\textwidth]{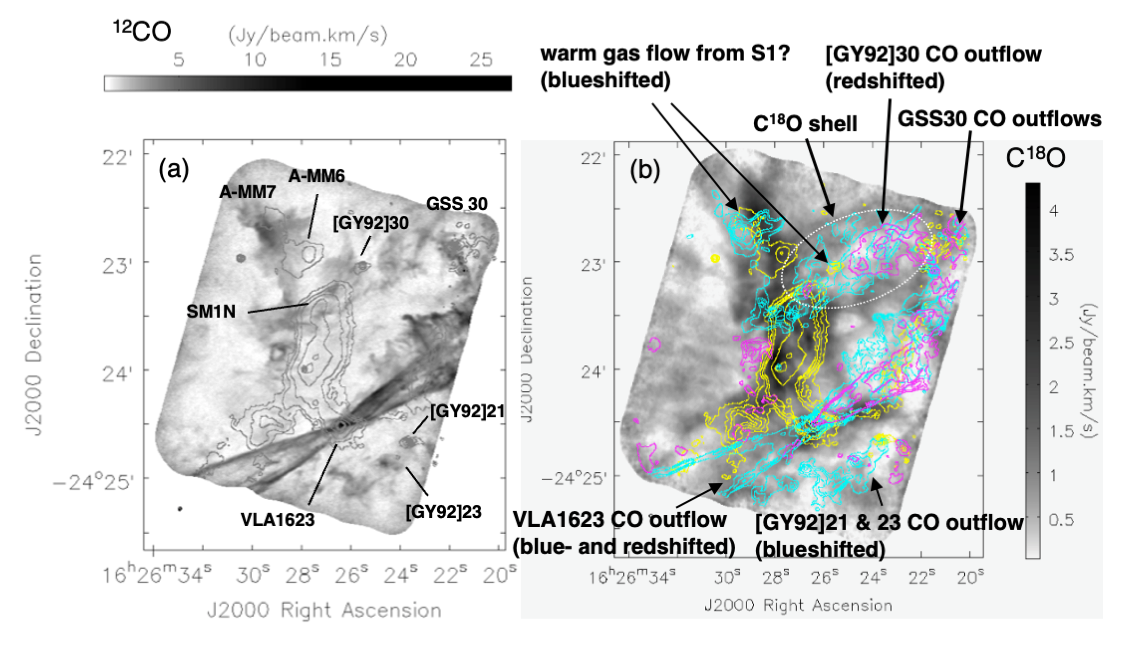}
\caption{\fn{
(a) $^{12}$CO ($J = 2$--$1$) and (b) C$^{18}$O ($J = 2$--$1$) velocity-integrated intensity images.  
In both panels, the gray and yellow contours are identical to those shown in Figure~\ref{fig:jwst}, for panels (a) and (b), respectively.  
In panel (b), we also show contours of the blueshifted (0.5 -- 1.5 km~s$^{-1}$; cyan) and redshifted (5.5 -- 6.5 km~s$^{-1}$; magenta) $^{12}$CO components. 
The both contours are drawn at 30, 60, 90, $\cdot$, 210$\sigma$ with 1 $\sigma$ 43 mJy beam$^{-1}$.
Several notable features are indicated with arrows (see the text).  
The expanding shell structure is highlighted by a dotted ellipse.
}}
    \label{fig:12COpeak}
\end{figure*}

Two primary formation scenarios are possible for these compact cores: turbulent fragmentation and ejection from a nearby stellar system like VLA 1623A/B.
In the first scenario, local cloud turbulence fragments the gas into cores, similar to how more massive cores form. If true, we would expect these cores to have velocities similar to those of the surrounding gas.
Such substellar-mass cores are reminiscent of other dust cores found in Oph A/B2 \citep{nakamura12b}.
In the second scenario, these effectively stable cores might be blobs created by gravitational instability within a circumstellar or circumbinary disk. Previous studies have suggested that such cores could be formed by the disintegration of multiple stellar systems \citep{reipurth01} or star-disk systems \citep{vorobyov10}, resulting in velocities distinct from the original stellar system’s systemic velocity. VLA 1623 ($V_{\rm LSR}\simeq 4$ km s$^{-1}$) is a strong candidate for such an ejector. Core 3, with its elongated shape, indeed may be interacting with the VLA 1623 outflow, which could explain its morphology. In contrast, other cores located outside the outflow lobe maintain a more rounded shape. 
Some weak CO finger-like structures are seen in the vicinity of the protostar (see the lowest panels of Figure \ref{fig:condensation}). These finger-like structures appear to be connected to the dust cores. 
Such structures might represent tracks created by the ejection process.
The actual structures may be affected by the interferometric missing flux.
Unfortunately, we could not find C$^{18}$O emission associated with these dust cores.
High-sensitivity, high-resolution molecular line observations would provide further insight into the kinematics of these cores, allowing us to constrain better their formation processes.
A similar scenario was discussed by \citet{murillo13}, \citet{harris18}, and \citet{mercimek23} on the formation of VLA 1623W.
As for the VLA 1623W, its systemic velocity of $0-1 $ km s$^{-1}$ \citep{murillo13} is somewhat different from VLA 1623A's velocity of $\sim 3.8 $ km s$^{-1}$ \citep{ohashi22}.
This velocity difference adds some credence to the ejection scenario.
A caveat is that it is uncertain that ejected cores can survive without significant distortion or even destruction due to hydrodynamical interaction with the surrounding gas.

Finally, we note that the dynamic interaction with protostellar outflow may also play a role in the formation and evolution of these cores and ALMA OphA 3, as discussed in the next subsection, might have evolved into a substellar object by the dynamic compression.

\subsubsection{Substellar Objects}
\label{subsub:brown drawf}

{\bf ALMA OphA 3}

As mentioned above, ALMA OphA 3 exhibits compact infrared emission detected in {\it Spitzer IRAC} at 4.5 $\mu$m and JWST at 4.7 $\mu$m. Its 1.3 mm peak emission is approximately 1.9 $\pm$ 0.12 mJy/beam, and its integrated intensity of 1.97 $\pm$ 0.21 mJy. Given an rms noise level of 0.25 mJy/beam, the peak emission corresponds to a significance of roughly 8 $\sigma$.

Figure \ref{fig:sed} shows the spectral energy distribution (SED) of this object, revealing an excess of near-infrared emission in the 1–10 $\mu$m range (black squares). 
For comparison, we also plotted the SEDs of a planetary-mass object OTS 44 \citep{Joergens13}, a young brown dwarf GY 204 (ISO-Oph 102 \citep{natta02,alves13}), and an extremely young Class 0 object ALMA Oph-2 with a mass of $\sim$ 0.1 M$_\odot$ \citep{kawabe18}. The luminosity of ALMA OphA 3 looks comparable to that of OTS 44, and is  weaker than that of GY 204. 
The low luminosity of ALMA OphA 3, $\approx$ a few $\times 10^{-3}$. L$_\odot$, classifies it as a Very Low Luminosity Object (VeLLO).
The envelope mass, derived from the 1.3 mm continuum emission, is estimated to be $3.53 \times 10^{-3}$ M$_\odot$. This low mass suggests that ALMA OphA 3 is a substellar object with an envelope or disk.
Even if all its envelope mass were to accrete onto the central object, its mass would be still in the substellar range.
Thus, we conclude that ALMA OphA 3 is a young brown dwarf or free-floating planetary-mass object candidate.
\fn{More complehensive SED analysis will be presented in a future papaer (Kawabe et al. 2025, in prep.)}
A slight elongation in the 1.3 mm dust continuum emission is observed (Figure \ref{fig:condensation}e and f), indicating dynamic interactions between the outflow and the surrounding dense gas. Similar such interactions could have potentially triggered the formation of the substellar object.

\begin{figure*}[htbp]
    \centering
\includegraphics[width=\textwidth]{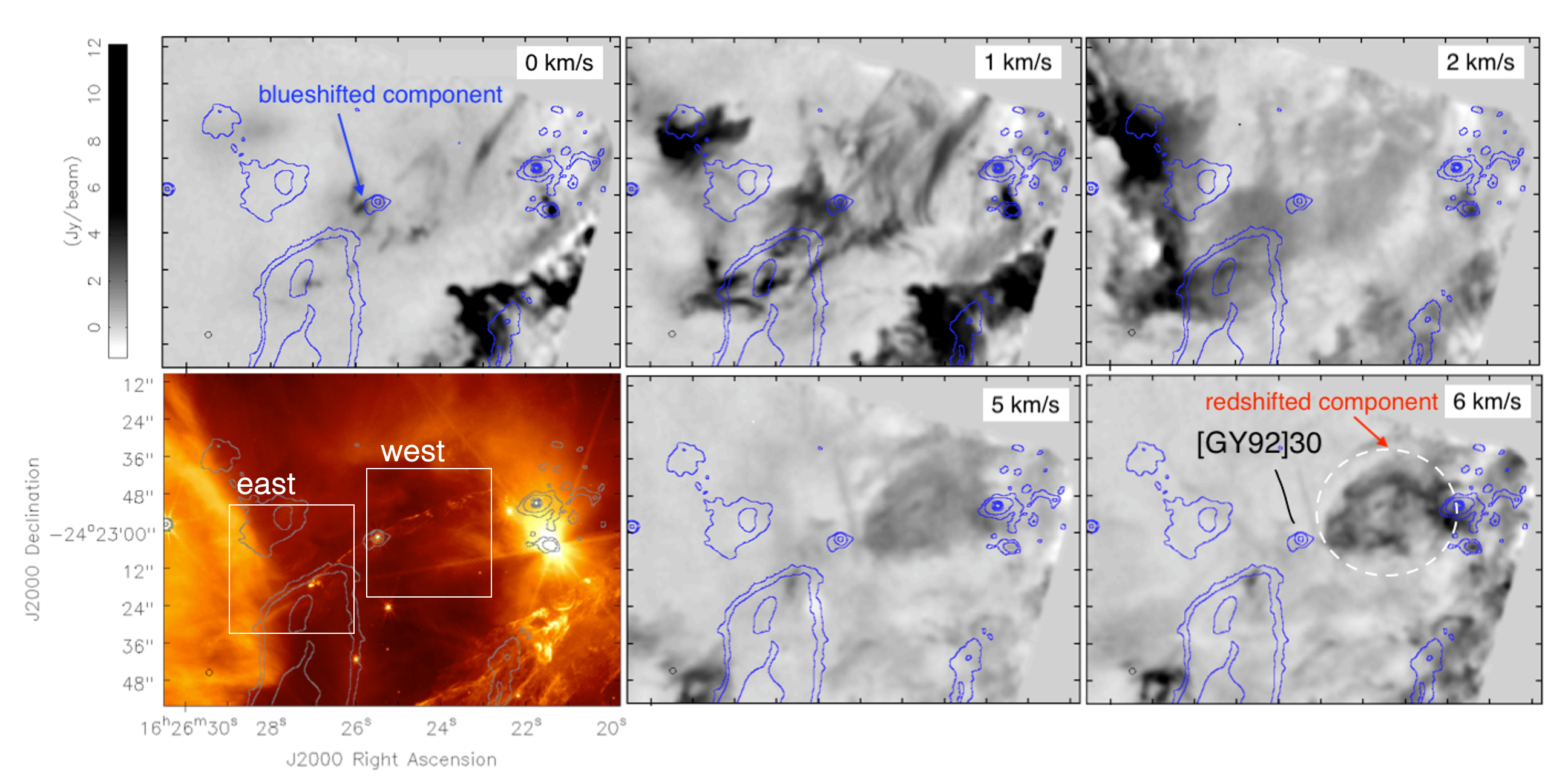}
    \caption{$^{12}$CO channel maps around the S1-GSS30 \fn{(from the northern part of the Oph A ridge to west)} region in the velocity range from 0 km s$^{-1}$ to 6 km s$^{-1}$. The 1.3 mm dust continuum emission is indicated in blue contours. 
    A dotted circle in the last panel indicates a part of the redshifted outflow component of [GY92]30, whereas the blueshifted component is shown in the first panel. In the lower-left panel, the JWST F470N image is presented and the protostellar jet from [GY92]30 is clearly seen. Two white boxes are the areas where $^{12}$CO and C$^{18}$O flux densities presented in Figure \ref{fig:coprofiles} are computed.} 
    \label{fig:channel_maps}
\end{figure*}

\begin{figure*}[htbp]
    \centering
\includegraphics[width=0.8 \textwidth]{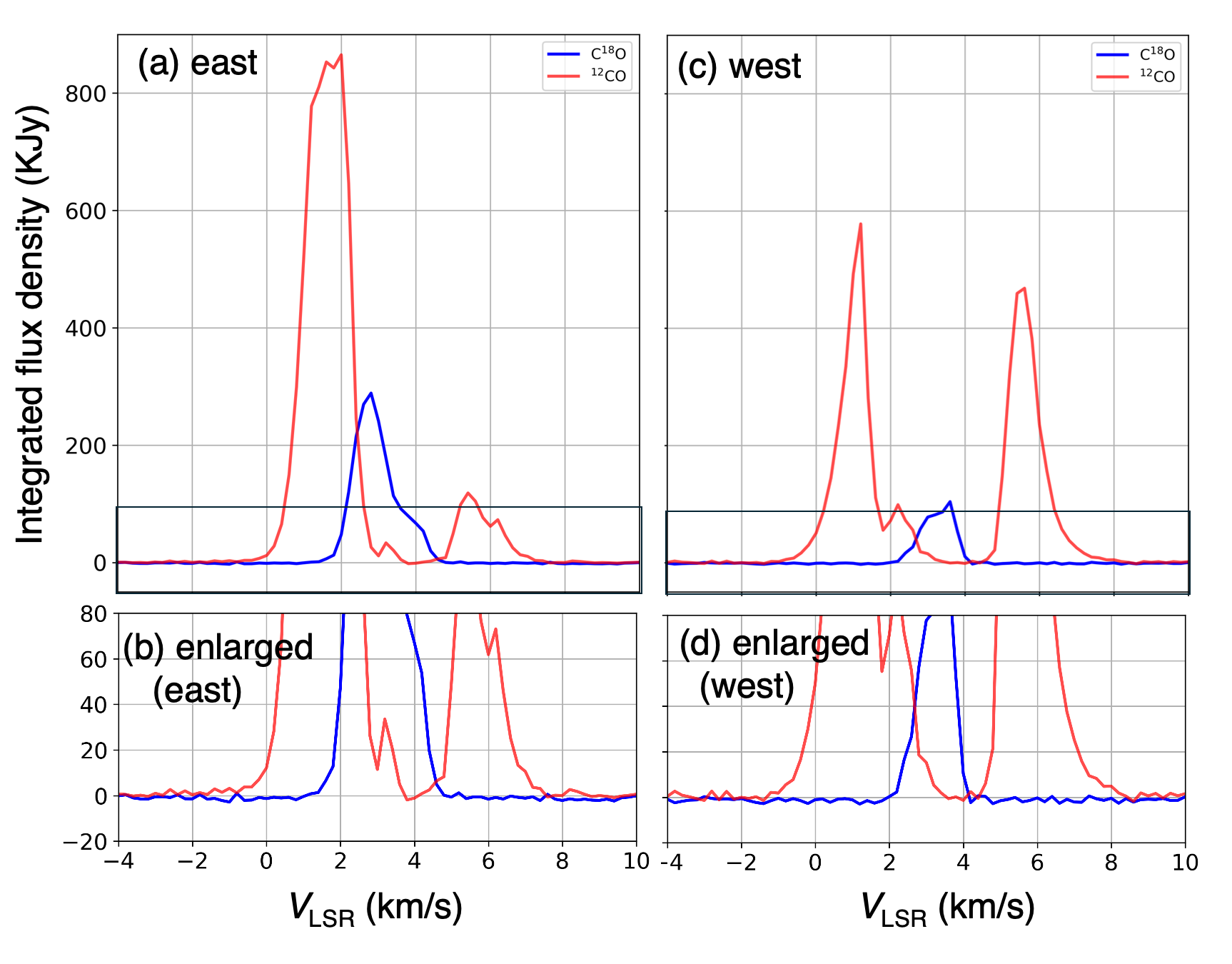}
    \caption{\fn{$^{12}$CO and C$^{18}$O integrated flux densities as a function of velocity at the east (a) and west (c) areas. The enlarged views of panels (a) and (c) are shown in panel (b) and (d). The red and blue colors represent the $^{12}$CO and C$^{18}$O profiles, respectively.} The areas taken are presented in Figure \ref{fig:channel_maps}. } 
    \label{fig:coprofiles}
\end{figure*}

{\bf ALMA OphA 4 and ALMA OphA 5}

From a comparison between JWST and ALMA 1.3 mm continuum images, we identified two faint cores. Both the 1.3 mm continuum emission and infrared emission from these cores are significantly weaker than those observed for ALMA OphA 3. This suggests that they are likely young substellar objects.  In particular, these cores are also detected at 4.5 $\mu$m in the {\it Spitzer} IRC band, with intensities approximately one-third that of ALMA OphA 3. 
Therefore, their luminosities are likely smaller than $10^{-3} L_\odot$. Their images are presented in Figures \ref{fig:condensation}.

\subsection{Stellar Feedback}

In a clustered environment, stellar feedback from forming stars, H{\sc ii} regions, PDRs, and protostellar outflows,  significantly influences subsequent star formation.
Here, we discuss the impact of the S1 expanding bubble (Section \ref{subsubsec:bubble}) and its complex structure (Section \ref{subsubsec:striations}), and then identify protostellar outflows with their driving sources (Section \ref{subsubsec:outflows}).

\subsubsection{Outflow of warm gas from the S1 bubble}
\label{subsubsec:bubble}

In Figure \ref{fig:12COpeak}, we show the $^{12}$CO \fn{and C$^{18}$O integrated} intensity maps.  We also present 
the $^{12}$CO channel maps of the S1-GSS30 area and averaged CO line profiles in Figures \ref{fig:channel_maps} \fn{and \ref{fig:coprofiles}, respectively}.\footnote{See also Figure 4 of \citet{hara21} for the $^{12}$CO channel maps of the southern part focusing on the VLA 1623.}
The relatively strong emission is concentrated along the eastern edge of the Oph A ridge, where a shell-like structure of $^{12}$CO emission is observed. 
This feature likely traces the PDR formed by the interaction between the dense ridge and the bubble driven by S1 (as seen prominently in the \fn{infrared} image).
The maximum $^{12}$CO peak intensity is located at the PDR shell with a brightness temperature ($T_{\rm mb}$) of 4.95 Jy/beam (= 89.6 K; see also \citet{yamagishi21} for the CI emission).
It is important to note that the $^{12}$CO maps are significantly affected by self-absorption and possibly by the resolved-out large-scale emission. Consequently, the observed peak temperature should be considered to be a lower limit of the actual gas kinetic temperature.

In the northern part of the ridge, the $^{12}$CO emission extends in an east-west direction, reaching toward the GSS30 protostar. Figure \ref{fig:channel_maps} shows this component is more prominent in the channel map of 1 km s$^{-1}$ and likely traces warm gas from the S1 shell, which appears to be blowing out between the A-MM6 and SM1N cores \fn{(see the blueshifted component in panel (b) of Figure \ref{fig:12COpeak}) and Figure \ref{fig:coprofiles}}. 
There is relatively strong $^{12}$CO emission located between A-MM6 and A-MM7. This component might also be from gas outflowing from the S1 bubble \fn{(see the blueshifted component in panel (b) of Figure \ref{fig:12COpeak})}.

Figure \ref{fig:12COpeak}(b) shows the $^{12}$CO emission extending toward GSS 30, which appears to be enclosed by C$^{18}$O emission.
This bubble-like structure might be also influenced by the redshifted lobe of the [GY92]30 outflow which is most prominent at the $^{12}$CO channel map of 5--6 km s$^{-1}$ (see the 5 km s$^{-1}$ and 6 km s$^{-1}$ channel images of Figure \ref{fig:channel_maps} and Section \ref{subsubsec:outflows}).
The redshifted component maintains a bowl-like shape, suggesting minimal interaction with the outflow gas (indicated by a dotted ellipse in the last panel of Figure \ref{fig:channel_maps}) whereas the blueshifted component appears very faint and less structured, hinting at potential interaction with the S1 warm outflow gas. 
The systemic velocity of S1 is measured to be --2.2 km s$^{-1}$ \citep{gutirrez20}, indicating that the $^{12}$CO gas is expanding at a velocity of about 3--4 km s$^{-1}$ relative to the star. With an assumed sound speed between 0.3 km s$^{-1}$ and 0.5 km s$^{-1}$, the estimated Mach number of the outflow reaches 6--10. If the outflow has a density of $10^{2-3}$ cm$^{-3}$, its dynamic pressure would be comparable to the thermal pressure of cold, dense gas with a density of $10^{4-5}$ cm$^{-3}$ at 10 K, suggesting that it likely contributes significantly to the formation of structure and the injection of momentum in this region.

\begin{figure*}[htbp]
    \centering
    \includegraphics[width= 1.02\textwidth]{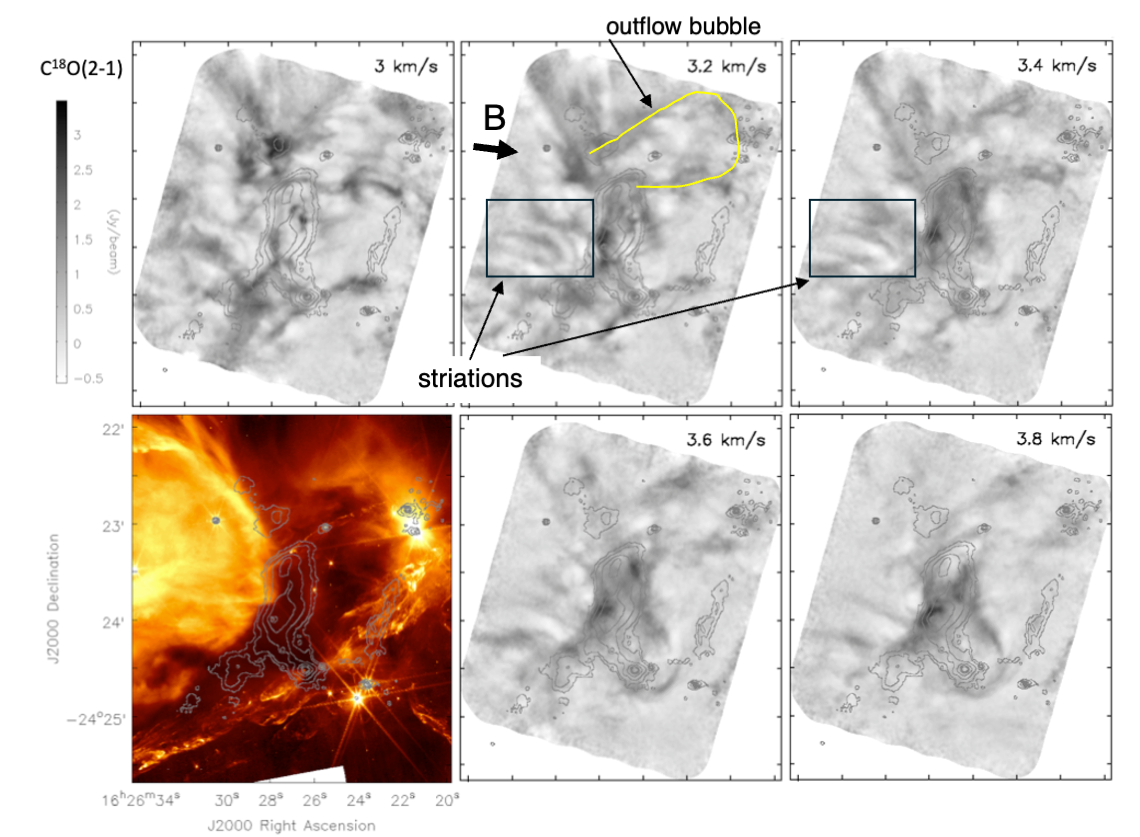}
    \caption{C$^{18}$O velocity channel maps (3.0 km s$^{-1}$ to 3.8 km s$^{-1}$) and striations in the S1 bubble.  Contours show the 1.3 mm dust continuum emission. Some striations are indicated in the boxes in the channel maps of 3.2 km s$^{-1}$ and 3.4 km s$^{-1}$ where the striations are prominent (see also Figures \ref{fig:c18opv} b and c).
    The magnetic field orientation in the bubble \citep{tram24} is indicated by an arrow in the panel of 3.2 km s$^{-1}$. For comparison, we also present the JWST F470N image with the 1.3 mm dust continuum emission and C$^{18}$O integrated intensity map in the bottom-left panel.}
    \label{fig:c18ochannelmap}
\end{figure*}

\subsubsection{C$^{18}$O Striations in the S1 bubble}
\label{subsubsec:striations}

In Figure \ref{fig:c18ochannelmap}, we display the C$^{18}$O channel maps ranging from 3 km s$^{-1}$ to 3.8 km s$^{-1}$. Interestingly, several filamentary structures are seen oriented in the east-west direction within the bubble. In addition,  a shell-like structure is seen extending toward GSS30 around the 3.2 km s$^{-1}$ channel (yellow dotted curve, see Section \ref{subsubsec:bubble}). Morphologically similar filamentary features were observed in the Oph B2 region by \citet{kamazaki19}. 

\begin{figure*}[htbp]
    \centering
\includegraphics[width=\textwidth]{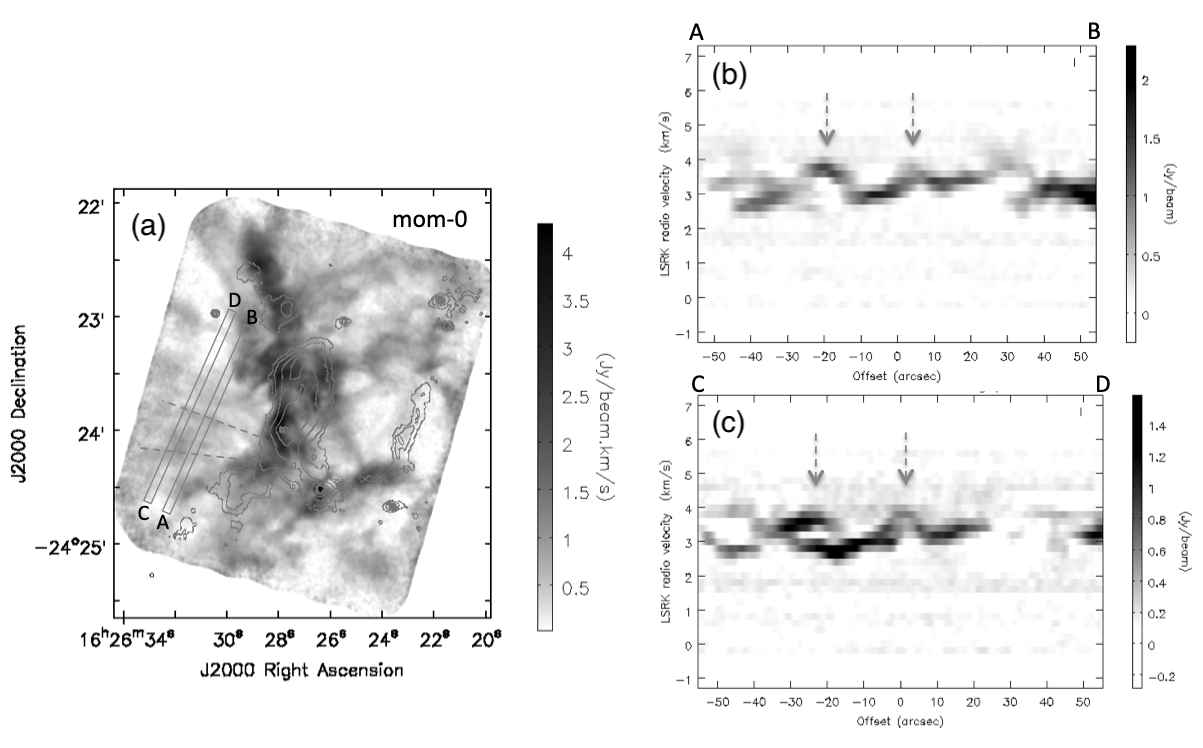}
    \caption{(a) C$^{18}$O moment-0 map with the two lines along which we took position-velocity maps, (b) and (c): C$^{18}$O position-velocity maps along the A--B and C--D lines indicated in panel (a), respectively.
The two dashed lines in panel (a) highlight the prominent striations, which are visible even in the integrated intensity image. These striations are most clearly observed in the 3.8 km s$^{-1}$ channel. The arrows in panels (b) and (c) point to the same striations.
 }
    \label{fig:c18opv}
\end{figure*}

An important characteristic of these structures is the alignment of the local magnetic field parallel to their long axes.
According to 154 $\mu$m continuum polarimetry \citep{tram24}, 
the magnetic field orientation between S1 and the Oph A ridge is predominantly in the east-west direction \citep[see also][for the magnetic field in the dense ridge]{kwon18}, with an estimated strength ranging from approximately 0.2 to 30 mG.
Interestingly, the magnetic field traced by near-infrared polarimetry over a much wider area shows a northeast-southwest alignment, which roughly coincides with the elongation of the S1 H{\sc ii}/PDR region \citep{kwon15}.
The plasma beta associated with the C$^{18}$O striations is measured to be 
\begin{equation}
\beta \equiv \frac{4\pi c_s^2 \rho}{B^2} \simeq 8.9 \times 10^{-3}  \ ,
\end{equation}
assuming a magnetic field strength of 0.2 mG (its lower limit), a temperature of 20 K, and a density of $10^4$ cm$^{-3}$ (or the Alfv\'en speed is evaluated to be about 3 km s$^{-1}$).
Thus, the shell is likely strongly magnetized. This indicates that the dynamics of the S1 H{\sc ii}/PDR bubble are largely controlled by the magnetic field.  
These filaments exhibit a quasi-periodic structure with separations of approximately 20--25\arcsec , ($\approx 3000-3500$ au, see Figures \ref{fig:c18opv}a/b/c). Figure \ref{fig:c18opv} shows position-velocity diagrams taken across the filaments, along the lines indicated in the left panel, revealing a velocity structure characterized by periodic undulations (the right two panels of Figure \ref{fig:c18opv}) that are reminiscent of magnetohydrodynamic (MHD) waves.

Comparable thin, quasi-periodic structures, often referred to as striations, have been observed in molecular cloud envelopes \citep{heyer08,skalidis23}. Two primary mechanisms are proposed for their formation given that the observed magnetic alignment suggests a significant role of magnetic fields. First, \citet{tritsis16} proposed that striations are the result of compressible magnetohydrodynamic (MHD) wave propagation \citep[see also][]{beattie20}. Second, striations may form due to the Kelvin-Helmholtz instability \citep{heyer16,hendrix15}, corrugations in magnetized sheetlike structure \citep{chen17}, or anisotropic turbulent phase mixing \citep{xu19}. 
Our C$^{18}$O striations are presumably substructures in the strongly magnetized bubble, and these  mechanisms can play a role in their formation.

\subsubsection{Protostellar Outflows}
\label{subsubsec:outflows}

Protostellar outflows and jets play a key role in driving supersonic turbulence within molecular clouds \citep{matzner07,nakamura11b,krumholz14, feddersen19}. Collimated or jet-like outflows, in particular, are highly efficient in injecting turbulence, as they create convective-like motions within bound clouds \citep{nakamura07,wang10}. 
Outflows from young protostars are typically detected through CO emission; however, as protostars evolve, CO emission becomes less prominent. Conversely, higher-velocity jet components are more effectively detected in infrared emission. Consequently, a combination of CO and infrared observations offers a promising method for identifying outflows and jets from more evolved protostars.

In this subsection, we identify protostellar outflows in the Oph A region using $^{12}$CO ($J=2-1$) observations and the JWST F470N image. Previous studies have identified molecular outflows in $\rho$ Oph through $^{12}$CO single-dish observations \citep{nakamura11, white15}. By integrating high-resolution $^{12}$CO data with the JWST infrared image, we not only confirmed previously known outflows but also identified new faint outflow candidates and their potential driving sources such as VLA 1623W, [GY92]21, and [GY92]23. \fn{In addition, we identified a bipolar jet from [GY92]20 which is located outside the ALMA mosaic area.} The newly identified driving sources of the outflows in Oph A are summarized in Table \ref{tab:objects}.

{\bf VLA1623 A/B:}

The second intensity peak in $^{12}$CO emission near VLA1623 reaches 4.73 Jy/beam ($\approx$ 85.7 K), clearly tracing the molecular outflow associated with VLA 1623 (see Figure \ref{fig:12COpeak}), as discussed in detail by \citet{hara21}. Their analysis provides insights into the density and velocity structures of this outflow.
See \citet{hara21} for more comprehensive discussion.

{\bf ALMA-OphA 1 and ALMA-OphA 2:}

\citet{kawabe18} discussed possible CO outflow components of ALMA-OphA 1 and ALMA-OphA 2
(See their Figures 1 and 6).
These objects are associated with very faint compact blueshifted CO components.

{\bf [GY92]30:}

In Figure \ref{fig:channel_maps}, we display velocity channel maps of the $^{12}$CO emission, covering the northern ridge region and extending to the area with the [GY92]30 and GSS30 protostars. The Class I object [GY92]30 shows an outflow with a prominent redshifted lobe extending in a bowl shape towards the west and overlapping with the GSS30 region at velocities of 5–7 km s$^{-1}$. This redshifted component retains its bowl-like shape, suggesting it is not directly interacting with the surrounding gas (particularly the outflow gas from S1) in 3D space. In contrast, the blueshifted component of [GY92]30 is compact and  localized at the source position around 0 km s$^{-1}$. 

{\bf GSS 30 IRS1/2/3:}

\begin{figure*}[htbp]
\includegraphics[width= \textwidth]{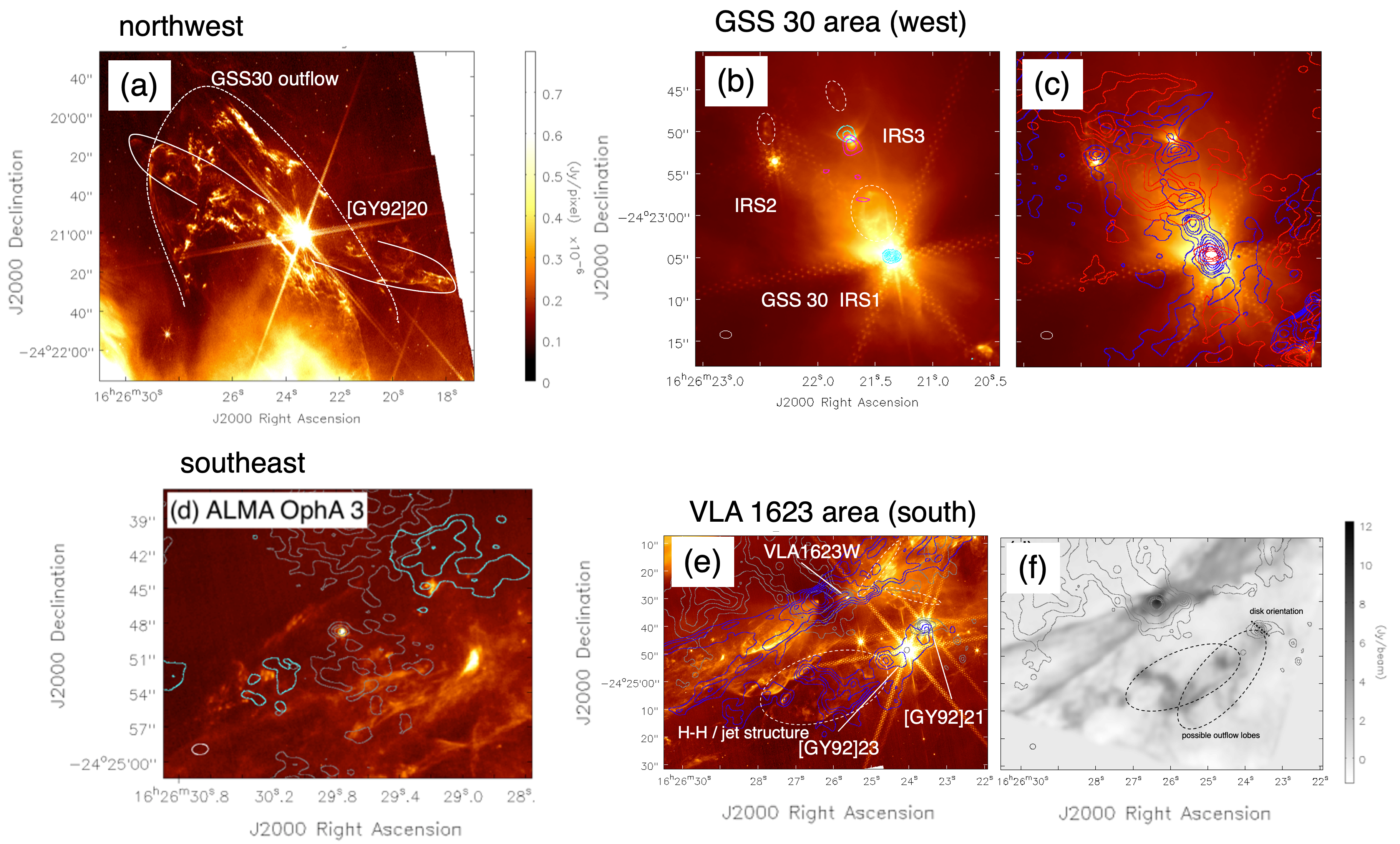}
\caption{
Zoomed-in views of several protostellar outflows or candidate sources in the Oph A region. The approximate locations (northwest, west, etc.) are labeled above each panel.
(a) Bipolar jet associated with [GY92] 20, traced by two solid curves in the JWST F470N image. The jet from GSS 30 extending northward is indicated by a dotted curve.
(b) Jets from GSS 30 IRS1/2/3 as seen in the F470N image, overlaid with high-velocity CO components. Cyan contours show CO integrated intensities in the velocity range --26.5 to --4.5 km,s$^{-1}$, and magenta contours represent the range --9.5 to 25.5 km,s$^{-1}$.
Contour levels for cyan are at (2.5, 5, 7.5, 10, 12.5, 15) $\times \sigma$, with $1\sigma = 0.234$ Jy beam$^{-1}$ km s$^{-1}$.
Magenta contours are at (2.5, 5) $\times \sigma$, with $1\sigma = 0.2$ Jy beam$^{-1}$ km s$^{-1}$.
(c) Same as panel (b), but showing low-velocity CO components from the GSS 30 IRS1/2/3 outflows. Blue contours integrate over --1 to 0 km,s$^{-1}$, and red contours cover 6.5 to 8.5 km,s$^{-1}$.
Both use contour levels of (6, 12, 18, 24, 36, 48, 60) $\times \sigma$, where $1\sigma = 0.07$ Jy beam$^{-1}$ km s$^{-1}$.
(d) Possible blueshifted CO component associated with ALMA OphA 3. Gray contours show the 1.3 mm dust continuum emission, drawn at 3, 5, 7 $\sigma$, etc.
(e) JWST F470N image overlaid with CO blueshifted components near VLA 1623W, [GY92] 21, and [GY92] 23.
Contours are at 3, 6, 9 $\sigma$, etc., with $1\sigma = 43$ mJy beam$^{-1}$ km s$^{-1}$.
(f) Blueshifted CO integrated intensity (2.1–2.5 km,s$^{-1}$) around VLA 1623W, [GY92] 21, and [GY92] 23.
Dashed ellipses mark the outflow components from [GY92] 21 and [GY92] 23. The dashed line shows the disk orientation of [GY92] 21.
}
    \label{fig:outflows}
\end{figure*}

Near GSS 30, three prominent outflow sources are recognized: GSS 30 IRS1, IRS2  and IRS3 (see Figure \ref{fig:channel_maps}). 
We present the spatial distributions of redshifted and blueshifted $^{12}$CO components around these sources in Figures \ref{fig:outflows}(a) and (b). As indicated in the Figure, all three objects seem to drive the outflows in a similar (north-south) direction. This orientation is consistent with the protostellar jet components indicated by dotted ellipses in Figure \ref{fig:jwst}(a).
GSS30 IRS3’s outflows are further discussed in recent work by \citet{santamaria24}.

{\bf VLA 1623W:}


Faint $^{12}$CO and infrared emission near VLA 1623W [Figure \ref{fig:outflows}(c) and (d)] suggests a potential outflow, despite the $^{12}$CO being weak and possibly contaminated by the nearby VLA 1623A/B outflows. The faint curved jet-like structures seen in the infrared, enclosed by two dotted ellipses in Figure \ref{fig:outflows}(c), suggest a weak jet from VLA 1623W.
Both structures look curved, and
can be almost connected perpendicularly to the flattened disklike envelope (its inclination angle $\approx 80^\circ$) found by \citep{mercimek23}. Therefore, we consider that either ellipse or both might be blowing out from this protostar.

{\bf [GY92]21 and [GY92]23 outflows:}

In Figure \ref{fig:outflows}(c) and (d), we can see molecular outflows and infrared jets from [GY92]21 and [GY92]23. A part of this CO high-velocity component was previously detected by \citet{kamazaki03} as AS outflow.
They suggested the possible driving source is [GY92]21 (LFAM3).  Our image indicates that each source appears to be associated with compact blueshifted CO components. 
The circumstellar disk of [GY92]21, whose orientation is indicated by a dotted line, is inclined with a position angle of about 45$^\circ$ \citep{cieza21}. Therefore, if the outflow is travelling in the direction perpendicular to the disk, it is likely to be in the dotted ellipse indicated in Figure \ref{fig:outflows}(d). 
The infrared components indicated by dotted ellipses in Figure \ref{fig:outflows}(c) are likely blowing out from [GY92]23, and the [GY92]23 outflow is nearly parallel to the VLA 1623 outflow, as indicated in Figure \ref{fig:outflows}.

\fn{{\bf [GY92]20 outflow:}}

\fn{This YSO lies outside the ALMA mosaic area, positioned near the northern bow-like structure of the GSS 30 outflow (see panel (a)).
Despite this, the JWST image clearly reveals its distinct bipolar jet components.
Based on their morphology, there is no indication of any dynamical interaction between the two outflows. Therefore, we interpret that they appear to be physically separate in 3D space.}

\fn{{\bf ISO-Oph 26 outflow:}}

\fn{This YSO lies outside the ALMA mosaic area, positioned in the southern outflow of GSS30. The collimated jet component is clearly seen in Figure \ref{fig:jwst}.
\citet{zhang13} is measured the proper motions of the knots in their Figure A.23. }

{\fn{\bf ALMA OphA 3 outflow?:}}

\fn{We detected weak blueshifted CO components around ALMA OphA3.
These components might be associated with an outflow from this object; however, they could also be part of the poweful VLA 1623 outflow.}

\section{Summary}
\label{sec:summary}

\fn{As the nearest cluster-forming region, the Ophiuchus molecular cloud complex, particularly the Oph A clump, offers an exceptional opportunity to study the earliest phases of star and substellar object formation. Its close proximity ($\sim$ 138 pc) provides a significant observational advantage, allowing us to resolve compact structures down to 10--10$^2$ au with currently available high-resolution and high-sensitivity instruments such as ALMA and JWST. This makes Oph A a uniquely well-suited benchmark region for testing theories of clustered star formation under Galactic physical conditions.}

\fn{By combining ALMA 7m+12m Array dust continuum and molecular line data with JWST infrared imaging, we have uncovered a wealth of structural details and signatures of active star formation. We identified several new faint compact substellar sources. Some of these sources are associated with infrared emission, indicating the presence of internal heating and reinforcing their identification as brown dwarfs or sub-brown dwarfs (free-floating planets). The detection of such extremely low-luminosity objects contributes to our understanding of the low-mass end of the stellar/substellar initial mass function (IMF), and raises important questions about the formation mechanisms of substellar objects in dense, clustered environments. These findings may provide observational constraints on whether brown dwarfs and sub-brown dwarfs form as scaled-down stars or via alternative pathways such as dynamical ejection or disk fragmentation.
In our analysis, some of the pre-substellar cores are located near the prototypical Class 0 object VLA 1623, and might have been ejected from this stellar system. In contrast, similar small cores were found along the possible outflow-interacting region at the western filament. Such interaction might promote fragmentation due to supersonic turbulence generated by outflow interaction. 
Revealing the complete populations of substellar-mass objects in this region may provide an important insight on the origin of substellar objects.}

\fn{Our data also reveal the complex impact of stellar feedback on the surrounding medium. The Herbig Be star S1 has carved out a bubble-like H{\sc ii} region with an associated PDR, whose influence extends across a substantial portion of the Oph A region. The Oph A ridge appears to lie along the surface of this bubble, and a break in the northern part of the ridge allows warm, low-density gas to flow outward toward the GSS 30 region. This escaping gas is confined laterally by intermediate-density structures traced by C$^{18}$O emission. Such structures may be responsible for momentum injection in molecular clouds \citep[e.g.,][]{krumholz14}. 
The Oph A ridge and western filament are located along the interacting regions of bubble and outflows. Such structures may suggest that the interaction driving by stellar feecback is important to create denser regions where future star formation occurs.
In addition, within the bubble interior, we detect parallel C$^{18}$O striations, whose linear morphology suggests they may arise from MHD wave in a strongly magnetized environment, potentially aligned with magnetic field lines. This highlights the critical interplay between magnetic fields, turbulence, and feedback in shaping the internal structure and kinematics of molecular clouds.}

\fn{Moreover, several new candidate protostellar outflows/jets have been identified by cross-analyzing ALMA and JWST data. One particularly striking feature is a bipolar infrared structure to the west, interpreted as a Herbig-Haro object, likely driven by an embedded protostar in the GSS 30 system, presumably GSS 30 IRS 1. In addition, we find that several compact objects near VLA 1623 are likely being dynamically influenced by its powerful outflow, suggesting that local feedback can have a strong effect even on individual core evolution and fragmentation.}

\fn{Taken together, our observations emphasize the pivotal role of feedback-both from intermediate-mass stars like S1 and from embedded protostars-in regulating the star formation process in clustered environments. Feedback influences the density distribution, drives turbulence, and may even trigger or suppress further star formation. In the context of the IMF, these mechanisms are likely to affect the core mass function and the eventual stellar mass distribution. Given that Ophiuchus is the closest region of clustered star formation accessible to current observatories, our results offer critical empirical input for theoretical models of molecular cloud evolution, star formation efficiency, and the origin of stellar and substellar populations.}

\fn{We believe that even a brief discussion of these broader implications enhances the context of our findings and underscores the relevance of Oph A as a template for understanding star formation in more distant and complex regions of the Galaxy.}

\clearpage

\acknowledgments

Data analysis was carried out on the multi-wavelength data analysis system operated by the Astronomy Data Center (ADC), National Astronomical Observatory of Japan.
This paper makes use of the following ALMA data: ADS/JAO.ALMA No. 2013.1.00839.S. ALMA is a partnership of ESO (representing its member states), NSF (USA) and NINS (Japan), together with NRC (Canada), MOST and ASIAA (Taiwan), and KASI (Republic of Korea), in cooperation with the Republic of Chile. The Joint ALMA Observatory is operated by ESO, AUI/NRAO and NAOJ. 
The National Radio Astronomy Observatory is a facility of the National Science Foundation operated under cooperative agreement by Associated Universities, Inc.
Some of the data presented in this paper were obtained from the Mikulski Archive for Space Telescopes (MAST) at the Space Telescope Science Institute. The specific observations analyzed can be accessed via \url{https://dx.doi.org/10.17909/hve0-q493}.
The Spitzer data were obtained from NASA/IPAC Infrared Science Archive page \citep{evans09} and the data can be accessed via
\url{https://dx.doi.org/10.1088/0067-0049/181/2/321}.
Fumitaka Nakamura was supported by the ALMA Japan Research Grant of NAOJ ALMA Project, NAOJ-ALMA-370.

\facility{ALMA, No:45m, Herschel}
\software{The data analysis in this paper uses python packages Astropy \citep{astropy13, astropy18, astropy22}, SciPy (Jones et al. 2001), Numpy (van der Walt et al. 2011), APLpy (Robitaille \& Bressert 2012), Matplotlib (Hunter 2007) and Astrodendro \citep{rosolowsky08}.}

\clearpage
\bibliography{nakamura}{}

\end{document}